\documentclass[a4paper,journal]{IEEEtran}

\makeatletter
\newcommand{\AddInputPath}[1]{%
  \ifx\input@path\@undefined
    \def\input@path{#1}
  \else
    \g@addto@macro{\input@path}{#1}
  \fi
}
\makeatother
\AddInputPath{{../}}

\usepackage{etex}

\usepackage[font=footnotesize,caption=false]{subfig} 

\usepackage{relsize}

\usepackage{array}
\usepackage{booktabs,tabularx}
\usepackage{multirow}
\usepackage[binary-units=true,per-mode=symbol,detect-mode=true]{siunitx}
\usepackage{graphicx}

\usepackage[T1]{fontenc}
\usepackage{textcomp}
\usepackage[utf8]{inputenc}
\usepackage[final]{microtype}
\usepackage{icomma}
\usepackage{xspace}

\usepackage[tbtags]{amsmath}
\usepackage{amssymb,amsfonts,bm}
\usepackage{mathtools} 
\usepackage{dsfont}
\usepackage{mathrsfs}
\usepackage{accents}
\usepackage{empheq}
\usepackage{nccmath}

\usepackage{color}
\usepackage{calc}
\usepackage{tikz}
\usepackage{pgfplots,pgfplotstable}
\usepackage{verbatim}
\usepackage{balance}

\usepackage[capitalize]{cleveref}
\usepackage{refcount}

\usepackage[inline,shortlabels]{enumitem}
\usepackage{algorithm}
\usepackage{algpseudocode}
\makeatletter
\newcommand{\algmargin}{\the\ALG@thistlm}
\makeatother
\newlength{\whilewidth}
\algdef{SE}[parWHILE]{parWhile}{EndparWhile}[1]
  {\parbox[t]{\dimexpr\linewidth-\algmargin}{%
     \hangindent\whilewidth\strut\algorithmicwhile\ #1\ \algorithmicdo\strut}}{\algorithmicend\ \algorithmicwhile}%
\algnewcommand{\parState}[1]{\State%
  \parbox[t]{\dimexpr\linewidth-\algmargin}{\strut #1\strut}}

\usepackage[acronym,shortcuts]{glossaries}
\usepackage{ifthen}
\usepackage{cite}
\usepackage{multibib}

\usepackage{comment}
\usepackage{todonotes}
\let\legacytodo\todo
\newcommand{\ruggedtodo}[2][]{\tikzexternaldisable\legacytodo[#1]{#2}\tikzexternalenable}
\renewcommand{\todo}[1]{\ruggedtodo[inline]{#1}}

\makeatletter
\def\todoref{\@ifnextchar[{\todoref@with}{\todoref@without}}
\def\todoref@without{\textbf{\color{red} [reference needed]}\xspace}
\def\todoref@with[#1]{\textbf{\color{red} [reference needed: #1]}\xspace}
\makeatother

\makeglossaries

\newacronym{fifo}{FIFO}{first in, first out}
\newacronym{dsl}{DSL}{digital subscriber line}
\newacronym{wsee}{WSEE}{weighted sum energy efficiency}
\newacronym{wpee}{WPEE}{weighted product energy efficiency}
\newacronym{wmee}{WMEE}{weighted minimum energy efficiency}
\newacronym{mmwave}{mmWave}{millimeter wave}
\newacronym{dfg}{DFG}{Deutsche Forschungsgemeinschaft}
\newacronym{haec}{HAEC}{Highly Adaptive Energy-Efficient Computing}
\newacronym{hpc}{HPC}{High Performance Computing}
\newacronym{mac}{MAC}{multiple-access channel}
\newacronym{bc}{BC}{broadcast channel}
\newacronym{siso}{SISO}{single-input/single-output}
\newacronym{simo}{SIMO}{single-input/multiple-output}
\newacronym{miso}{MISO}{multiple-input/single-output}
\newacronym{mimo}{MIMO}{multiple-input/multiple-output}
\newacronym{af}{AF}{amplify-and-forward}
\newacronym{df}{DF}{decode-and-forward}
\newacronym{cf}{CF}{compress-and-forward}
\newacronym{mwrc}{MWRC}{multi-way relay channel}
\newacronym{dmmwrc}{DM-MWRC}{discrete memoryless multi-way relay channel}
\newacronym{pde}{PDE}{partial data exchange}
\newacronym{fde}{FDE}{full data exchange}
\newacronym{iid}{i.i.d.\@}{independent and identically distributed}
\newacronym{dm}{DM}{difference-of-monotonic}
\newacronym{dc}{DC}{difference-of-convex}
\newacronym{awgn}{AWGN}{additive white Gaussian noise}
\newacronym{awg}{AWG}{additive white Gaussian}
\newacronym{sic}{SIC}{successive interference cancellation}
\newacronym{snr}{SNR}{signal-to-noise ratio}
\newacronym{sinr}{SINR}{signal-to-interference-plus-noise ratio}
\newacronym{inr}{INR}{interference to noise ratio}
\newacronym{zf}{ZF}{zero-forcing}
\newacronym{mrt}{MRT}{maximum ratio transmission}
\newacronym{mmse}{MMSE}{minimum mean square error}
\newacronym{sud}{SUD}{single user decoding}
\newacronym{dof}{DoF}{degrees of freedom}
\newacronym{gdof}{GDoF}{generalized degrees of freedom}
\newacronym{nnc}{NNC}{noisy network coding}
\newacronym{dmn}{DMN}{discrete memoryless network}
\newacronym{csi}{CSI}{channel state information}
\newacronym{pmf}{pmf}{probability mass function}
\newacronym{dmic}{DM-IC}{discrete memoryless interference channel}
\newacronym{ic}{IC}{interference channel}
\newacronym{ee}{EE}{energy efficiency}
\newacronym{gee}{GEE}{global energy efficiency}
\newacronym{tin}{TIN}{treating interference as noise}
\newacronym{snd}{SND}{simultaneous non-unique decoding}
\newacronym{sd}{SD}{simultaneous decoding}
\newacronym{hk}{HK}{Han-Kobayashi}
\newacronym{rs}{RS}{rate splitting}
\newacronym{rf}{RF}{radio frequency}
\newacronym{lna}{LNA}{low noise amplifier}
\newacronym{lo}{LO}{local oscillator}
\newacronym{adc}{ADC}{analog-to-digital converter}
\newacronym{dac}{DAC}{digital-to-analog converter}
\newacronym{dsp}{DSP}{digital signal processing}
\newacronym{brd}{BRD}{best response dynamics}
\newacronym{br}{BR}{best response}
\newacronym{ne}{NE}{Nash equilibrium}
\newacronym{lhs}{LHS}{left-hand side}
\newacronym{rhs}{RHS}{right-hand side}
\newacronym{ran}{RAN}{radio access network}
\newacronym{qos}{QoS}{Quality of Service}
\newacronym{ngmn}{NGMN}{Next Generation Mobile Networks}
\newacronym{cap}{CAP}{Capacity Adaptation}
\newacronym{bwa}{BW}{Bandwidth Adaptation}
\newacronym{prb}{PRB}{physical resource block}
\newacronym{se}{SE}{spectral efficiency}
\newacronym{tp}{TP}{throughput}
\newacronym{bs}{BS}{base station}
\newacronym{mop}{MOP}{multi-objective optimization problem}
\newacronym{gda}{GDA}{generalized Dinkelbach's algorithm}
\newacronym{midcp}{MIDCP}{mixed integer disciplined convex programming}
\newacronym{lp}{LP}{linear program}
\newacronym{brb}{BRB}{branch-reduce-and-bound}
\newacronym{bb}{BB}{branch-and-bound}
\newacronym{pa}{PA}{Polyblock Algorithm}
\newacronym{sit}{SIT}{successive incumbent transcending}
\newacronym{oma}{OMA}{orthogonal multiple access}
\newacronym{noma}{NOMA}{non-orthogonal multiple access}
\newacronym{wlog}{w.l.o.g.\@}{without loss of generality}
\newacronym{lsc}{l.s.c.\@}{lower semi-continuous}
\newacronym{usc}{u.s.c.\@}{upper semi-continuous}
\newacronym{mm}{MM}{mixed monotonic}
\newacronym{mmp}{MMP}{mixed monotonic programming}

\glsenableentrycount
\makeglossaries

\usetikzlibrary{positioning}
\usetikzlibrary{calc}
\usetikzlibrary{fit}

\usetikzlibrary{external}

\crefname{equation}{}{}
\crefrangeformat{equation}{(#3#1#4)--(#5#2#6)}
\crefformat{footnote}{#2\footnotemark[#1]#3}

\newcommand{\abs}[1]{\ensuremath{\left\lvert #1 \right\rvert}}

\DeclareMathOperator\diam{diam}

\DeclareMathOperator*{\argmax}{arg\,max}
\DeclareMathOperator*{\argmin}{arg\,min}

\let\vec\bm

\DeclareMathOperator{\st}{s.t.}
\newcommand{\zero}{\bm{0}}

\newcommand{\brbF}{F}
\newcommand{\brbf}{f}
\newcommand{\brbG}{G}

\newcommand{\cpeps}{\varepsilon_\mt{CP}}
\newcommand{\asmeps}{\varepsilon_\mt{tol}}

\newcommand{\brbBox}[2]{\left[{#1},~{#2}\right]}
\newcommand{\brbbox}[2]{[{#1}, {#2}]}

\newcommand{\brbxsymb}{x}
\newcommand{\brbysymb}{y}
\newcommand{\brbasymb}{r}
\newcommand{\brbbsymb}{s}

\newcommand{\brbx}{\bm\brbxsymb}
\newcommand{\brbxk}[1][\brbk]{\brbxsymb_{#1}}
\newcommand{\brby}{\bm\brbysymb}
\newcommand{\brbyk}[1][\brbk]{\brbysymb_{#1}}
\newcommand{\brba}{\bm\brbasymb}

\newcommand{\brbak}[1][\brbk]{\brbasymb_{#1}}

\newcommand{\brbb}{\bm\brbbsymb}

\newcommand{\brbbk}[1][\brbk]{\brbbsymb_{#1}}

\newcommand{\mb}[1]{\bm{#1}}
\newcommand{\mt}[1]{\mathrm{#1}}
\newcommand{\Tr}{{\operatorname{T}}}
\newcommand{\He}{{\operatorname{H}}}
\newcommand{\one}{\boldsymbol{1}} 
\newcommand{\id}{\mathbf{I}} 

\newcommand{\noise}{\eta}
\newcommand{\notk}{j}
\newcommand{\allk}{{\forall k}}
\newcommand{\Ms}{L}
\newcommand{\ms}{\ell}

\newcommand{\pow}{p}

\allowdisplaybreaks[1]

\newtheorem{theorem}{Theorem}
\newtheorem{lemma}{Lemma}
\newtheorem{corollary}{Corollary}
\newtheorem{proposition}{Proposition}

\newtheorem{definition}{Definition}
\newtheorem{XXXremark}{Remark}
\newenvironment{remark}
	{\begin{XXXremark}}
	{\xqed{\lozenge}\end{XXXremark}}

\newcommand{\xqed}[1]{%
	\leavevmode\unskip\penalty9999 \hbox{}\nobreak\hfill
	\quad\hbox{\ensuremath{#1}}}

\newtheorem{XXXassumption}{Case}

\newtheorem{XXXexample}{Example}

\newcounter{optimizationproblem}

\newenvironment{optprob*}{\begin{equation*}\left\{\begin{aligned}}{\end{aligned}\right.\end{equation*}\ignorespacesafterend}

\newcommand{\nofracX}[2]{{#1}\,{#2}^{-1}}

\pgfplotscreateplotcyclelist{default}{%
	blue,mark=*\\%
	red,mark=triangle*\\%
	teal,mark=square*\\%
	brown!60!black,mark=diamond*\\%
	cyan,every mark/.append style={solid},mark=star\\%
	black,every mark/.append style={solid,fill=lightgray},mark=otimes*\\%
}

\allowdisplaybreaks[2]

\newcommand\chg[1]{#1}

\begin{document}
\title{Mixed Monotonic Programming\\for Fast Global Optimization}

\author{Bho~Matthiesen,~\IEEEmembership{Member,~IEEE},
		Christoph~Hellings,~\IEEEmembership{Member,~IEEE},\\
		Eduard~A.~Jorswieck,~\IEEEmembership{Fellow,~IEEE},
		and~Wolfgang~Utschick,~\IEEEmembership{Senior~Member,~IEEE}%
		\thanks{
			B.\,Matthiesen \chg{is with the Department of Communications Engineering, University of Bremen, 28359 Bremen, Germany (e-mail: matthiesen@uni-bremen.de).}
			\chg{C.\,Hellings is with ETH Z\"urich, Department of Physics, 8093 Z\"urich, Switzerland (e-mail: chellings@phys.ethz.ch).}
			E.\,A.\,Jorswieck is with Technische Universität Braunschweig, Institut für Nachrichtentechnik, 38106 Braunschweig, Germany (e-mail: jorswieck@ifn.ing.tu-bs.de).
			\chg{W.\,Utschick is with Technische Universit\"at M\"unchen, Professur f\"ur Methoden der Signalverarbeitung, 80290 M\"unchen, Germany (e-mail: utschick@tum.de).}
			Part of this research was conducted while \chg{B.\,Matthiesen and E.\,A.\,Jorswieck were} with Technische Universität Dresden and \chg{C.\,Hellings was with Technische Universit\"at M\"unchen}.
		}%
		\thanks{
			This work is supported in part by the German Research Foundation (DFG) in the
		Collaborative Research Center 912 ``Highly Adaptive Energy-Efficient Computing,'' \chg{under Germany's Excellence Strategy (EXC 2077 at University of Bremen, University Allowance),} and under grant number JO~801/24-1.
			We thank the Center for Information Services and High Performance Computing (ZIH) at TU
			Dresden for generous allocations of computer time.
			\chg{B. Matthiesen and C. Hellings contributed equally to this work.}
		}%
		\thanks{
			Conference versions of two application examples can be found in \cite{MaHeJo19,HeMaJoUt19}. This journal version gives a more general perspective on the framework, gives more details on algorithmic aspects, and discusses many further application examples.
		}%
	}

\maketitle

\begin{abstract}
While globally optimal solutions to \chg{many} convex programs can be computed efficiently in polynomial time,
this is, in general, not possible for nonconvex optimization problems.
Therefore, locally optimal approaches or other efficient suboptimal heuristics are usually applied for practical implementations.
However, there is also a strong interest in computing globally optimal solutions of nonconvex problems in offline simulations in order to benchmark the faster suboptimal algorithms.
Global solutions often rely on monotonicity properties.
A common approach is to reformulate problems into \chg{a canonical monotonic} optimization problem where the monotonicity becomes evident,
but this often comes at the cost of nested optimizations, increased numbers of variables, and/or slow convergence.
\chg{The framework of mixed monotonic programming (MMP) proposed in this paper avoids such performance-deteriorating reformulations by revealing hidden monotonicity properties directly in the original problem formulation.}
By means of a wide range of application examples from the area of signal processing for communications
(including energy efficiency for green communications, resource allocation in interference networks, scheduling for fairness and quality of service, as well as beamformer design in multiantenna systems),
we demonstrate that the novel MMP approach leads to tremendous complexity reductions compared to state-of-the-art methods for global optimization.
\chg{However, the framework is not limited to optimizing communication systems, and we expect that similar speed-ups can be obtained for optimization problems from other areas of research as well.}

\end{abstract}
\glsresetall


\section{Introduction}
\label{sec:intro}
In point-to-point communication systems without interference, the optimization of various performance metrics can be formulated as convex programs \chg{such as in rate maximization \cite{Te99} or mean square error minimization \cite{PaCiLa03}}.
More complicated objective functions in the context of energy efficiency optimization \chg{can be} shown to be pseudoconvex or quasiconvex \cite{Isheden2012}.
Even in advanced scenarios \chg{with} multiple antennas or parallel transmission on orthogonal carriers,
these optimization problems can be solved with efficient methods from convex optimization \cite{PaCiLa03} or fractional programming \cite{Isheden2012}, and sometimes even in closed form \cite{Te99}.
However, in multi-terminal scenarios with interfering users, performance optimization typically involves nonconvex problems.
This is \chg{often} due to interference terms that make the rate equations nonconcave or due to product operations contained in multiuser utility functions.

Apart from special cases where efficient solutions
exist,\footnote{E.g.,
in \cgls{mimo} broadcast channels with dirty paper coding \cite{Jindal2005}
or for rate balancing problems in \cgls{miso} interference channels with interference treated as noise \cite{Liu2012}.}
performance optimization in interference networks is, thus, usually tackled by locally optimal approaches or suboptimal heuristics.
Examples are gradient ascent algorithms 
\cite{YeBl03,BoKa08,HuScJo08,HeUtJo11},
successive allocation methods
\cite{YoGo06,GuUtHuJo10},
successive \mbox{(pseudo-)}convex approximation
\cite{Zappone2016,Yang2017},
alternating optimization
\cite{CoToJuLa07,ShScBo08,ChAgCaCi08,LaAgVi16},
distributed interference pricing
\cite{ScShBeHoUt09},
or game-theoretic methods 
\cite{LaJo08,ScPaBa08a,Matthiesen2015,Baccelli2011}.
Such heuristics are good candidates for practical implementation due to their low computational complexity and/or the possibility of distributed implementation.
However, \chg{there is also a strong interest in globally optimal solutions} to assess the fundamental \chg{limits} of the considered multiuser communication systems and to have benchmarks for the heuristic methods.

In order to obtain such global solutions, researchers have applied methods from the field of \chg{monotonic optimization \cite{Horst1996,Tuy2000,Tuy2005,Tuy2016}} to optimization problems in various communication systems.
For instance, monotonic programming was applied in interference channels \cite{JoLa08,Jorswieck2010a,Qian2009,Qian2010,Zhang2012,Liu2012,Zappone2015,Zappone2017,UtBr12},
in broadcast channels with linear transceivers \cite{Br12,BrUt09,HeUtJo11,HeJoRiUt12,HeUt12},
in interfering broadcast channels \cite{Bjoernson2012},
in relaying scenarios \cite{Matthiesen2015,Zappone2015},
and in satellite systems \cite{GrJoUt12}
with the aim of maximizing weighted sum rates \cite{JoLa08,Jorswieck2010a,Qian2009,Zhang2012,Liu2012},
fairness-based performance metrics \cite{Qian2010,Zhang2012,BrUt09,Br12,UtBr12,Bjoernson2012},
or the energy efficiency \cite{HeUt12,Matthiesen2015,Zappone2017,Zappone2015}
as well as minimizing the required sum transmit power \cite{HeUtJo11,HeJoRiUt12,GrJoUt12}.
Some of these applications include solutions for multiantenna systems \cite{JoLa08,Jorswieck2010a,Liu2012,BrUt09,Br12,UtBr12,Bjoernson2012,HeUtJo11,HeJoRiUt12},
allow to average data rates over several time slots \cite{Qian2010,Zhang2012,UtBr12,Br12,HeJoRiUt12},
and/or incorporate additional robustness considerations \cite{Bjoernson2012}.
Moreover, monotonic optimization can also be applied \chg{on the medium access control layer. One example is optimizing the} transmit probabilities in the slotted ALOHA protocol \cite{Zhang2012}.
A wider overview with further application examples can be found in \cite{Zhang2012,Bjornson2013,Zappone2015}.

A common approach is to reformulate the objective function\footnote{Similar reformulations can be applied to the constraints if needed.} as a difference
$ f^+(\vec x) - f^-(\vec x)$  of nondecreasing functions $f^+$ and $f^-$.
The resulting \cgls{dm} problem can \chg{be reformulated further into} a canonical monotonic optimization problem
where a nondecreasing function is maximized over a normal set.\footnote{A set $\mathcal{G}\subset\mathbb{R}_0^+$ is called normal if $[\zero; \mb x]\subseteq\mathcal{G}$ for all $\mb x\in\mathcal{G}$ \cite{Tuy2000}.}
For instance, instead of maximizing $ f^+(\vec x) - f^-(\vec x)$ over a box $[\brba, \brbb]$,
we can maximize the nondecreasing function $f^+(\vec x) + t$ under the additional constraints $f^-(\vec x) +t \leq 0$ and $t\in[-f^-(\brbb), -f^-(\brba)]$ which \chg{form} a normal set.\footnote{Please refer to \cite[Thm.~11.1]{Tuy2016} for more details.}
The resulting canonical monotonic optimization problem can then be solved with the so-called \cgls{pa} \cite[Sec.~11.2]{Tuy2016} as was done, e.g., in \cite{JoLa08,Jorswieck2010a}.
An important drawback is that the number of optimization variables is increased by introducing the auxiliary variable $t$. \chg{This negatively} affects the convergence speed \chg{because the \cgls{pa} has exponential worst-case complexity in} the number of variables \cite{Bjornson2013}.

As an alternative, \cgls{dm} problems can be solved by means of \chg{\cgls{bb}} techniques as described in \cite{Tuy2005}.
This approach, which was \chg{pursued in} \cite{GrJoUt12,HeUtJo11,HeJoRiUt12,Bjoernson2012}, avoids the overhead of the additional optimization variable $t$,
but still suffers from \chg{drawbacks that will be observed} in \cref{sec:appl:wsrmax}.
Just like the \cgls{pa}, \chg{\cgls{bb}} methods rely on calculating utopian bounds to the objective function,
and \chg{their} convergence speed \chg{depends heavily} on the quality of these bounds.
Unfortunately, \cgls{dm} bounds are, \chg{in general,} not very \chg{tight.}

Therefore, several authors have proposed to \chg{improve the speed of convergence by reparameterizing an optimization problem in terms of a new set of monotonic variables.}
For instance, \cite{Qian2009,Qian2010,Zhang2012} use the \cgls{sinr} values of the users as optimization variables instead of their transmit powers,
while \cite{Liu2012,BrUt09,Br12,UtBr12,HeUtJo11,HeJoRiUt12} use the achievable rates, and \cite{GrJoUt12} uses the received interference powers.
The resulting monotonic or \cgls{dm} problems can then be solved by means of the \cgls{pa} \cite{Qian2009,Qian2010,Zhang2012,BrUt09,Br12,UtBr12} or a \chg{\cgls{bb}} algorithm \cite{GrJoUt12,HeUtJo11,HeJoRiUt12}.
However, the change of variables usually makes the evaluation of the objective and constraint functions more costly.
For instance, the \cgls{sinr} values and achievable rates can be calculated analytically when the transmit powers are used as optimization variables,
but \chg{a fixed point iteration} is necessary to calculate the transmit powers if the \cgls{sinr} values or the achievable rates are used as variables (see, \cite{Qian2009,Qian2010,Zhang2012} and \cite{Liu2012,BrUt09,Br12,UtBr12,HeUtJo11,HeJoRiUt12}, respectively).
\chg{Thus, a} change of variables might reduce the number of iterations required in the monotonic programming method, but \chg{comes} at the cost of increasing the computational complexity of each iteration.

Moreover, not all optimization problems can be conveniently rewritten in terms of monotonic functions or \cgls{dm} functions.
For instance, in the context of energy-efficient communications,
we encounter objective functions that can be written as fractions of \cgls{dm} functions.
For this type of problems, the fractional monotonic programming method proposed in \cite{Matthiesen2015,Zappone2015,Zappone2017} uses a monotonic programming approach as an inner solver inside Dinkelbach's method for fractional programs.
This combination has the drawback that a highly complex monotonic programming algorithm has to be executed not only once but repeatedly in each iteration of the outer algorithm.
Moreover, it is no longer possible to obtain a rigorous guarantee that the obtained solution is indeed $\eta$-optimal, i.e., that it is no more than a given constant $\eta$ away from the exact globally optimal solution.

In this paper, we propose the framework of \cgls{mmp} which avoids all these drawbacks since it neither requires a reformulation of the objective function nor a change of variables.
Instead, the main idea is that a function defined by several terms might have different monotonicity properties in each term and variable.
Thus, the \cgls{mmp} approach does not consider whether the whole function is monotonic in a variable, but takes the monotonicity for each occurrence of a variable separately into account
by formulating a so-called \cgls{mm} function.
If such an \cgls{mm} function can be constructed for a given optimization problem, the problem can be solved by \chg{a \cgls{bb} algorithm} as discussed in \cref{sec:algo}.\footnote{\chg{A \cgls{brb} algorithm is a special kind of \cgls{bb} algorithm that includes a reduction step to speed up the convergence.}}
In \cref{sec:appl}, we show that a wide variety of optimization problems (including the difficult fractional monotonic problems mentioned above)
can be solved with the \cgls{mmp} approach, and we demonstrate significant advantages compared to state-of-the-art solutions
using the C++ implementation available at \cite{github}.

Note that there are several existing approaches that can be considered as special cases of the \cgls{mmp} framework, the most prominent being \cgls{dm} formulations.
However, the \cgls{mmp} approach is much more general and can be used to find solution methods that are faster than the \cgls{dm} approach.
This will become clear after the formal definition of an \cgls{mmp} problem in \cref{sec:prob}.
Moreover, some specialized solution methods developed for particular optimization problems can be identified to fall into the more general \cgls{mmp} framework.
For instance, \cite{HeUt12} exploited a structure with a fraction of nonnegative nondecreasing functions of a scalar variable,
and \cite{HeUt18,HeUt19} consider an optimization \chg{problem} in a two-user interference channel that can be identified \chg{as} a two-dimensional special case of the \cgls{mmp} framework.
An implementation of the \cgls{brb} algorithm for \cgls{mmp} problems can, thus, be readily applied to any of these special cases.

\emph{Notation:}
We use $\zero$ for the zero vector, $\one$ for the all-ones vector, and $\id_L$ for the identity matrix of size $L$. 
Vectors are written in bold-face lowercase \chg{and} matrices in bold-face uppercase.
Inequalities between vectors are meant component-wise, \chg{i.e., $\vec x \ge \vec y$ if and only if $x_i \ge y_i$ for all $i$,} and $\brbBox{\brba}{\brbb}=\{\brbx \,|\, \brba\leq\brbx\leq\brbb\}$ denotes a box (hyperrectangle).
We use shorthand notations of the form $(\bullet_k)_\allk=(\bullet_1,\dots,\bullet_K)$,
and we write $\mathcal{CN}(0, 1)$ for the circularly symmetric Gaussian distribution with zero mean and unit variance.

\section{Mixed Monotonic Programming}
\label{sec:prob}
Consider the optimization problem
\begin{equation} \label{eq:MMP}
	\underset{\vec x\in\mathcal D}{\text{max}}\enskip f(\vec x) \tag{P}
\end{equation}
with continuous objective function \chg{$f : \mathds R^n \to \mathds R$} and compact feasible set $\mathcal D\subseteq\mathds R^n$.
For now, we do not need any further assumptions on $\mathcal D$ and postpone the discussion of its structure to \cref{sec:feasibility}.
Let $\mathcal M_0 = [\vec r^0, \vec s^0]$ be a box 
in $\mathds R^n$ enclosing $\mathcal D$, i.e., $\mathcal M_0 \supseteq \mathcal D$.
Assume there exists a continuous function \chg{$F : \mathds R^n \times \mathds R^n \to \mathds R$} such that
\begin{equation}
	F(\vec x, \vec x) = f(\vec x)
	\label{mmp:prop3}
\end{equation}
for all $\vec x\in\mathcal M_0$ and
\begin{subequations}
	\label{mmp:prop12}
\begin{align}
	\brbF(\brbx,\brby) &\leq \brbF(\brbx',\brby)\qquad \text{if}\ \brbx\leq\brbx', \label{mmp:prop1}\\
	\brbF(\brbx,\brby) &\geq \brbF(\brbx,\brby')\qquad \text{if}\ \brby\leq\brby'. \label{mmp:prop2}
\end{align}
\end{subequations}
for all $\vec x, \vec x'\!, \vec y, \vec y' \in \mathcal M_0$.
We call such a function a \glsreset{mm}\emph{\cgls{mm}} function. The optimization problem \cref{eq:MMP} is said to be a \glsreset{mmp}\emph{\cgls{mmp}} problem if its objective has an \cgls{mmp} representation, i.e., if $f$ satisfies \cref{mmp:prop3} for some \cgls{mm} function $F$.
In the following section, we will show that \cgls{mmp} problems are especially well suited for solution by a \chg{\cgls{bb}} procedure.

As mentioned before, some well established problem formulations can be identified as special cases of \chg{this} novel \cgls{mmp} framework.
The most prominent among them are \cgls{dm} programs \cite{Tuy2000}, i.e.,
\begin{equation} \label{opt:monotonic}
	\underset{\vec x\in\mathcal D}{\text{max}}\enskip f^+(\vec x) - f^-(\vec x)
\end{equation}
where $f^+$ and $f^-$ are nondecreasing functions. A \cgls{mmp} representation of that objective is $F(\vec x, \vec y) = f^+(\vec x) - f^-(\vec y)$.

However, the \cgls{mmp} approach is much more versatile. For example, consider the fraction
\begin{equation} \label{eq:frac}
	\frac{p^+(\vec x) - p^-(\vec x)}{q(\vec x)}
\end{equation}
with nondecreasing $p^+$, $p^-$, and $q$,
where we assume 
$p^+(\vec x) - p^-(\vec x)\geq0$ and $q(\vec x)>0$ for all $\vec x$.
Maximizing this function with monotonic programming requires the combination of Dinkelbach's algorithm \cite{dinkelbach1967} as outer and monotonic programming as inner solver \cite{Zappone2017}. This approach has the drawbacks that the inner global optimization problem needs to be solved several times and the stopping criterion does not guarantee an $\eta$-optimal solution. Instead, 
\chg{\cref{eq:frac} can be optimized directly by the algorithm proposed in \cref{sec:algo} since it is easily verified that
\begin{equation}
	F(\vec x, \vec y) = \frac{p^+(\vec x) - p^-(\vec y)}{q(\vec y)}
\end{equation}
is an \cgls{mmp} representation of \cref{eq:frac}.}

\chg{It is important to note that the \cgls{mmp} representation of $f$ is never unique. This can be observed in the following simple example. Let $F$ be an \cgls{mmp} representation of $f$. Then, it is easy to verify that
	\begin{equation} \label{eq:badmmp}
	\tilde F(\vec x, \vec y) = F(\vec x, \vec y) + \sum_{i=1}^N (x_i - y_i)
\end{equation}
fulfills the requirements in \cref{mmp:prop12,mmp:prop3} as well. Intuitively, $\tilde F$ can be understood as an \cgls{mmp} version of
\begin{equation}
	\tilde f(\vec x) = f(\vec x) + \sum_{i=1}^N (x_i - x_i).
\end{equation}
However, while we obviously have $\tilde f(\vec x)=f(\vec x)$, the difference between the MMP representations $\tilde F$ and $F$ is crucial. As we will see later, $F$ leads to tighter bounds than $\tilde F$ which, in turn, leads to faster convergence of the \cgls{bb} algorithm. 

Another, practically more relevant, example for the non-uniqueness of $F$ is throughput maximization in
wireless interference networks \cite{Zappone2017}}
\begin{equation} \label{opt:sr}
	\underset{\vec 0 \le \vec p \le \vec P}{\text{max}}\enskip \sum_{i=1}^K \log\left( 1+ \frac{\alpha_i p_i}{\sigma^2_i + \sum_{j=1}^K \beta_{i,j} p_j} \right)
\end{equation}
with positive constants $\alpha_i$, $\sigma_i$, and nonnegative $\beta_{i,j}$. Conventionally, \cref{opt:sr} is converted into a \cgls{dm} program \cref{opt:monotonic} with $f^+(\vec x) = \sum_{i=1}^K \log( \alpha_i x_i + \sigma^2_i + \sum_{j = 1}^K \beta_{i,j} x_j )$ and $f^-(\vec x) = \sum_{i=1}^K \log( \sigma^2_i + \sum_{j = 1}^K \beta_{i,j} x_j )$. This yields the \cgls{mmp} representation in the text below \eqref{opt:monotonic}. A more direct approach to obtain $F$ from \cref{opt:sr} is
\begin{equation}
	F(\vec x, \vec y) = \sum_{i=1}^K \log\left( 1+ \frac{\alpha_i x_i}{\sigma^2_i + \beta_{i,i} x_i + \sum_{j \neq i} \beta_{i,j} y_j} \right).
\end{equation}
\chg{This example will be continued in \cref{sec:appl:wsrmax}. An important aspect discussed there is how the precise choice of $F$ directly impacts the convergence speed of the developed algorithm.}

To conclude this section, we state some useful properties of \cgls{mm} functions.
\chg{Let $F_i(\vec x, \vec y)$ be \cgls{mm} functions for $i = 1, \dots, K$. Then,
\begin{gather} \label{eq:mmrules:sum}
	(\vec x, \vec y) \mapsto \sum_{i = 1}^K F_i(\vec x, \vec y),
	\\
	(\vec x, \vec y) \mapsto \!\max_{i = 1, \ldots, K} F_i(\vec x,\vec y),\hspace{.8em}
(\vec x, \vec y) \mapsto \!\min_{i = 1, \ldots, K} F_i(\vec x,\vec y) \label{eq:mmrules:maxmin}
\end{gather}
are \cgls{mm} functions, i.e., the properties of \cgls{mm} functions are preserved by summation and by taking the pointwise minimum or maximum of several \cgls{mm} functions.
Moreover, if $g(\vec x)$ is a real-valued, nondecreasing function, and $h(\vec x)$ is a real-valued, nonincreasing function, the composed functions
\begin{align}
	(\vec x, \vec y) &\mapsto g(F_i(\vec x, \vec y)), &
	(\vec x, \vec y) &\mapsto h(F_i(\vec y, \vec x)) \label{eq:mmrules:incdec}
\end{align}
are \cgls{mm} functions as well.
Note that the nondecreasing and nonincresing variables are swapped in case of a composition with a nonincreasing function, i.e.,
to ensure that the composition of $h$ and $F_i$ is nondecreasing in its first argument and nonincreasing in the second one, the first argument of the composition has to be plugged into $F_i$ as the second argument and vice versa.}
In particular, it follows from \cref{eq:mmrules:incdec} that $(\vec x, \vec y) \mapsto -F(\vec y, \vec x)$ and $(\vec x, \vec y) \mapsto 1/F(\vec y, \vec x)$ are \cgls{mm} if $F(\vec x, \vec y)$ is a positive \cgls{mm} function.
If in addition $F_i(\vec x, \vec y) \ge 0$ for all $i = 1, \dots, K$ and $\vec x, \vec y \in \mathcal X$ for some $\mathcal X \subseteq \mathds R^n$, then
\begin{equation}
	(\vec x, \vec y) \mapsto \prod\nolimits_{i = 1}^{K} F_i(\vec x,\vec y) \label{eq:mmrules:prod}
\end{equation}
is an \cgls{mm} function on $\mathcal X$, \chg{i.e., the product of nonnegative \cgls{mm} functions is \cgls{mm} as well.}

\section{Global Optimal Solution of \eqref{eq:MMP}}
\label{sec:algo}
We design a \chg{\cgls{brb}} algorithm to determine a global $\eta$-optimal solution of \cref{eq:MMP}, i.e., a feasible point $\bar{\vec x}\in\mathcal D$ such that
$f(\bar{\vec x}) \ge f(\vec x) - \eta$
for all $\vec x\in\mathcal D$.
The core idea of \chg{any \cgls{bb} algorithm, including the considered \cgls{brb} variant,} is to relax the feasible set $\mathcal D$ and subsequently partition it such that upper bounds on the objective value can be determined easily.\footnote{\chg{Please refer to \cite[Chapter 3]{diss}, \cite[Chapter 4]{Horst1996}, or \cite[Chapter 6]{Tuy2016} for a thourough introduction to \cgls{bb} methods.}} This is where the \cgls{mmp} representation $F$ of the objective function $f$ comes in handy as it is
well suited to compute upper bounds over rectangular sets.
Let $\mathcal M = [\vec r, \vec s]$
be a box in $\mathds R^n$. Then,
\begin{equation}
\label{eq:algo:mmbound}
	\max_{\vec x\in\mathcal M\cap\mathcal D} f(\vec x) \le \max_{\vec x\in\mathcal M} F(\vec x, \vec x) \le \max_{\vec x, \vec y\in\mathcal M} F(\vec x, \vec y) = F(\vec s, \vec r)
\end{equation}
gives an upper bound
$U(\mathcal M)=U([\vec r,\vec s])=F(\vec s, \vec r)$
on the optimal value of $f(\vec x)$ on $\mathcal M\cap\mathcal D$.
Thus, rectangular subdivision \cite[Sec.~6.1.3]{Tuy2016}, where a box $\mathcal M$ is partitioned along a hyperplane parallel to one of its facets, is an excellent choice to partition $\mathcal D$.
Given a point $\vec v\in\mathcal M$ and index $j \in\{ 1, 2, \dots, n\}$, we divide $\mathcal M$ along the hyperplane $x_j = v_j$. The resulting partition sets are the subrectangles
\begin{subequations}
\label{eq:partition}
\begin{align}
	\mathcal M^- &= \{ \vec x  \,|\, r_j \le x_j \le v_j,\ r_i \le x_i \le s_i\ (i\neq j) \} \\
	\mathcal M^+ &= \{ \vec x \,|\, v_j \le x_j \le s_j,\ r_i \le x_i \le s_i\ (i\neq j) \}.
\end{align}
\end{subequations}
This is referred to as a \emph{partition via $(\vec v, j)$} of $\mathcal M$.
A partition of $\mathcal M$ via $(\frac{1}{2} (\vec s + \vec r), j)$ where $j \in\argmax_j s_j - r_j$ is called a \emph{bisection} of $\mathcal M$.

We say that $\{ \mathcal M_k \}$ is a \emph{decreasing sequence of sets} if, for all $k$, $\mathcal M_{k+1} \subset \mathcal M_k$, i.e., $\mathcal M_{k+1}$ is a descendent of $\mathcal M_k$.
The following proposition is an important property for the convergence of \cgls{bb} methods.
\begin{lemma}[\kern-.82ex{\cite[Corollary~6.2]{Tuy2016}}] \label{prop:exhaustive}
	Let $\{\mathcal M_k\}$ be a decreasing sequence of sets such that $\mathcal M_{k+1}$ is a descendent of $\mathcal M_k$ in a bisection \chg{along a longest side of $\mathcal M_k$}. Then, the diameter $\diam(\mathcal M_k)$ of $\mathcal M_k$ tends to zero as $k \rightarrow\infty$.
\end{lemma}

Besides the subdivision procedure and computation of bounds, the selection of the next box (or branch) for further partitioning is crucial for the convergence and implementation of \chg{a} \cgls{bb} procedure. A widely used selection criterion is
\begin{equation}
	\mathcal M_k \in \argmax\{ U(\mathcal M) \,|\, \mathcal M\in\mathscr R_{k-1} \}.
	\label{eq:selection:argmax}
\end{equation}
where $U(\mathcal M)$ is the upper bound chosen for the \chg{\cgls{bb}} method --- in our case the \cgls{mmp} bound defined below \eqref{eq:algo:mmbound} --- and
$\mathscr R_{k-1}$ holds all undecided boxes from the previous \chg{iteration \cite[Sec.~6.2]{Tuy2016}}, i.e., all boxes for which it is not yet clear whether or not they contain the global optimum.
However, this selection might not be the best choice from an implementation point of view. To guarantee convergence, it suffices if the selection satisfies the following condition.
\begin{definition}[\kern-.82ex{\cite[Def~IV.6]{Horst1996}}] \label{def:selection}
	A selection operation is said to be \emph{bound improving} if, at least each time after a finite number of steps, $\mathcal M_k$ satisfies \cref{eq:selection:argmax}.
\end{definition}
By construction, \cref{eq:selection:argmax} satisfies \cref{def:selection}.
An alternative is to select one of the oldest elements in $\mathscr R_{k-1}$, i.e., define for every $\mathcal M$ by $\sigma(\mathcal M)$ the iteration index of its creation and select
\begin{equation}
	\mathcal M_k \in \argmin\{ \sigma(\mathcal M) \,|\, \mathcal M\in\mathscr R_{k-1} \}.
	\label{eq:selection:oldest}
\end{equation}
Due to the finiteness of $\mathscr R_k$ every set $\mathcal M\in\mathscr R_k$ will be deleted or selected after finitely many iterations \cite[p.~130]{Horst1996}.

The final \cgls{brb} procedure is stated in \cref{alg:bb}.
\chg{It is initialized in \ref{alg:bb:init} where an initial box $\mathcal M_0 = [\vec r^0, \vec s^0]$ containing the feasible set $\mathcal D$ is required, i.e.,
\begin{align}
	r^0_i &\le \min_{\vec x\in\mathcal D} x_i & s^0_i &\ge \max_{\vec x\in\mathcal D} x_i
\end{align}
for all $i = 1, 2, \dots, n$. In \ref{alg:bb:branch}, a box is selected for partitioning with any bound improving selection rule, e.g., \cref{eq:selection:argmax} or \cref{eq:selection:oldest}, and then bisected along one of its longest dimensions. \ref{alg:bb:reduction} is optional and discussed separately in \cref{sec:appl:red}. For each newly constructed box, a feasible value is computed in \ref{alg:bb:update}. If necessary, the current best known feasible solution $\bar{\vec x}^k$ (the ``incumbent'') and current best known value $\vec \gamma_k$ are updated. Infeasible boxes, i.e., new boxes that do not contain any feasible points, are deleted (pruned) in \ref{alg:bb:prune}. Note that the box selected in \ref{alg:bb:branch} is replaced by the new boxes and, thus, removed from the partition $\mathscr R_k$. In \ref{alg:bb:termination}, the algorithm is terminated if the partition is empty or if none of the remaining boxes can contain any better solution. Otherwise, the algorithm continues in \ref{alg:bb:branch}.}
\begin{algorithm}[htb]
	\caption{\cGls{brb} Algorithm for \cgls{mmp} Problems} \label{alg:bb}
	\small
	\centering
	\begin{minipage}{\linewidth-1em}
		\begin{enumerate}[label=\textbf{Step \arabic*},ref=Step~\arabic*,start=0,leftmargin=*]
			\item\label{alg:bb:init} {\bfseries (Initialization)} Choose $\mathcal M_0 \supseteq \mathcal D$ and $\eta > 0$. Let $k=1$ and $\mathscr R_0 = \{\mathcal M_0\}$. If available or easily computable, find $\bar{\vec x}^0\in\mathcal D$ and set $\gamma_0 = f(\bar{\vec x}^0)$. Otherwise, set $\gamma_0 = -\infty$.
			\item\label{alg:bb:branch} {\bfseries (Branching)}
				Select a box $\mathcal M_k = [\vec r^k, \vec s^k] \in \mathscr R_{k-1}$ and bisect $\mathcal M_k$ via $(\frac{1}{2} (\vec s^k + \vec r^k), j)$ with $j \in\argmax_j s_j^k - r_j^k$. Let $\mathscr P_{k} = \{ \mathcal M^-_k, \mathcal M^+_k\}$ with $\mathcal M^-_k$, $\mathcal M^+_k$ as in \cref{eq:partition}.
			\item\label{alg:bb:reduction} {\bfseries (Reduction)}
				For each $\mathcal M\in\mathscr P_k$, replace $\mathcal M$ by $\mathcal M'$ such that $\mathcal M'\subseteq\mathcal M$ and
				\begin{equation}\label{eq:alg:bb:reduction}
					(\mathcal M\setminus\mathcal M') \cap \{\vec x\in\mathcal D \,|\, F(\vec x, \vec x) > \gamma_k \} = \emptyset.
				\end{equation}
			\item\label{alg:bb:update}  {\bfseries (Incumbent)}
			For each $\mathcal M\in\mathscr P_k$, find $\vec x \in\mathcal M\cap\mathcal D$ and set $\alpha(\mathcal M) = f(\vec x)$. If $\mathcal M\cap\mathcal D = \emptyset$, set $\alpha(\mathcal M) = -\infty$.
				Let $\alpha_k = \max\{ \alpha(\mathcal M) \,|\, \mathcal M \in\mathscr P_k\}$. If $\alpha_k > \gamma_{k-1}$, set $\gamma_k = \alpha_k$ and let $\bar{\vec x}^k\in\mathcal D$ such that $\alpha_k = f(\bar{\vec x}^k)$. Otherwise, let $\gamma_k = \gamma_{k-1}$ and $\bar{\vec x}^k = \bar{\vec x}^{k-1}$.
			\item\label{alg:bb:prune}  {\bfseries (Pruning)}
				Delete every $\mathcal M = [\vec r, \vec s]\in\mathscr P_k$ with $\mathcal M\cap\mathcal D = \emptyset$ or $F(\vec s, \vec r) \le \gamma_k + \eta$. Let $\mathscr P_k'$ be the collection of remaining sets and set $\mathscr R_{k} = \mathscr P_k' \cup (\mathscr R_{k-1}\setminus\{ \mathcal M_k \})$.
			\item\label{alg:bb:termination}  {\bfseries (Termination)}
				Terminate if $\mathscr R_{k} = \emptyset$ or, optionally, if $\{[\vec r, \vec s]\in\mathscr R_k \,|\, F(\vec s, \vec r) > \gamma_k + \eta \} = \emptyset$. Return $\bar{\vec x}^k$ as a global $\eta$-optimal solution. Otherwise, update $k \leftarrow k + 1$ and return to Step 1.
		\end{enumerate}
	\end{minipage}
\end{algorithm}
\chg{Convergence of \cref{alg:bb} to an $\eta$-optimal solution} \chg{of \cref{eq:MMP} is established below\footnote{We combine the convergence proof from \cite[Prop.~5.6]{Tuy1998}, \cite[Prop.~6.1]{Tuy2016} with the idea of a general selection criterion from \cite[Thm.~IV.3]{Horst1996}.} for any $F$ satisfying \cref{mmp:prop12,mmp:prop3}.}
\begin{theorem} \label{thm:bbconv}
	\Cref{alg:bb} converges towards a global $\eta$-optimal solution of \cref{eq:MMP} if the selection is bound improving.
\end{theorem}

\begin{IEEEproof}
	In \ref{alg:bb:reduction}, let $\mathcal D' = \{\vec x\in\mathcal D \,|\, F(\vec x, \vec x) > \gamma_k \} \subseteq \mathcal D$ and observe that $\mathcal D\setminus\mathcal D'$ does not contain any solutions better than the current best solution. Thus, if $\mathcal M'$ satisfies \cref{eq:alg:bb:reduction}, no solutions better than the current incumbent are lost and the reduction does not affect the solution of \cref{eq:MMP}.

	If the algorithm terminates in \ref{alg:bb:termination} and iteration $K$, then $F(\vec s, \vec r) \le \gamma_K + \eta$ for all $\mathcal [\vec r, \vec s] \in\mathscr R_K$ and, since $F(\vec x, \vec y) > -\infty$, $\gamma_K > -\infty$. Hence, $\bar{\vec x}_K$ is feasible and
	\begin{equation} \label{thm:bbconv:max}
		\gamma_K = f(\bar{\vec x}_K) \ge F(\vec r, \vec s) - \eta \ge \max_{\vec x\in\mathcal M\cap \mathcal D} f(\vec x) -\eta
	\end{equation}
	for every $\mathcal M \in \mathscr R_K$. Now, for every $\vec x\in\mathcal D$, either $\vec x \in\mathcal D\cap \bigcup_{\mathcal M\in\mathscr R_K} \mathcal M$ or $\vec x \in\mathcal D\setminus \bigcup_{\mathcal M\in\mathscr R_K} \mathcal M$.
	In the first case, $f(\vec x) -\eta \le f(\bar{\vec x}_K)$ due to \cref{thm:bbconv:max}.
	In the latter case, $\vec x\in\mathcal M'$ for some $\mathcal M'=[\vec r', \vec s']\in\mathscr P_{k'}\setminus \mathscr P_{k'}'$ and some $k'$ since $\mathcal M_0\supseteq\mathcal D$. Because $\{ \gamma_k \}$ is nondecreasing and due to \ref{alg:bb:prune},
		$f(\vec x) \le F(\vec s', \vec r') \le \gamma_{k'} + \eta \le \gamma_K + \eta$.
	Hence, for every $\vec x\in\mathcal D$, $f(\vec x) \le f(\bar{\vec x}_K) + \eta$.

	It remains to show that \cref{alg:bb} is finite. Suppose this is not the case. Then, due to the bound improving selection, there exists an infinite decreasing subsequence of sets $\{ \mathcal M_{k_q} \}_q$ such that $\mathcal M_{k_q} \in \argmax\{F(\vec s, \vec r) \,|\, [\vec r, \vec s]\in\mathscr R_{k_q}\}$.
	Because $\mathcal M_{k_q} \cap \mathcal D \neq \emptyset$, there exists an $\vec x^{k_q}\in\mathcal M_{k_q}\cap\mathcal D$.
	Due to \cref{prop:exhaustive}, $\diam\mathcal M_{k_q} \rightarrow 0$ as $q\rightarrow \infty$. Thus, $\vec x^{k_q}$, $\vec s^{k_q}$, and $\vec r^{k_q}$ all converge towards a common limit, and, together with \cref{mmp:prop3},
	$F(\vec s^{k_q}, \vec r^{k_q}) \rightarrow F(\vec x^{k_q}, \vec x^{k_q}) = f(\vec x^{k_q})$.
	Since $\alpha([\vec r, \vec s]) \le \sup_{\vec x\in[\vec r, \vec s]\cap \mathcal D} f(\vec x) \le F(\vec s, \vec r) \le F(\vec s^{k_q}, \vec r^{k_q})$ for all $\mathcal [\vec r, \vec s] \in\mathscr R_{k_q}$, $F(\vec s^{k_q}, \vec r^{k_q}) \rightarrow \gamma_{k_q}$ and, hence,
	$F(\vec s^{k_q}, \vec r^{k_q}) = \gamma_{k_q} + \delta_{k_q}$ with $\delta_{k_q} \ge 0$ and $\lim_{q\rightarrow\infty}\delta_{k_q} = 0$.
	Thus, there exists a $\tilde K$ such that $\delta_{k} \le \eta$ for all $k > \tilde K$, and $F(\vec s, \vec r) \le \gamma_{k} + \eta$ for all $[\vec r, \vec s]\in\mathscr R_{k}$ and $k > \tilde K$. Then, either the algorithm is directly terminated in \ref{alg:bb:termination} or the remaining sets in $\mathscr R_k$ are successively pruned in finitely many iterations until $\mathscr R_k = \emptyset$.
\end{IEEEproof}

\begin{remark}[Relative Tolerance]
	\Cref{alg:bb} determines an $\eta$-optimal solution of \cref{eq:MMP}, i.e., a feasible solution $\bar{\vec x}$ of \cref{eq:MMP} that satisfies $f(\bar{\vec x}) \ge f(\vec x) - \eta$ for all $\vec x\in\mathcal D$. Instead, by replacing all occurrences of ``$\gamma_k + \eta$'' in \cref{alg:bb} with ``$(1+\eta) \gamma_k$'', the tolerance $\eta$ becomes relative to the optimal value and the algorithm terminates if the solution satisfies $(1+\eta) f(\bar{\vec x}) \ge  f(\vec x)$ for all $\vec x\in\mathcal D$. The necessary modifications of \cref{thm:bbconv} are straightforward.
\end{remark}

\chg{\begin{remark}[Non-Uniqueness of \cgls{mmp} Representations]
The proof of \cref{thm:bbconv} is valid for any $F$ satisfying \cref{mmp:prop12,mmp:prop3}. 
Thus, the non-uniqueness of the \cgls{mmp} representation $F$ does not impair the convergence proof of \cref{alg:bb}.
However, the actual choice of $F$ has an impact on the tightness of the obtained bounds and, thus, on the convergence speed.
For the example in \cref{eq:badmmp}, we can calculate $\tilde F(\vec s, \vec r) - F(\vec s, \vec r) = \sum_{i=1}^N (s_i - r_i) \geq 0$
to see that the MMP representation $F$ never leads to worse bounds than the alternative $\tilde F$.
A practical example in which the influence of the choice of the \cgls{mmp} representation on the convergence speed can be observed is studied in detail in \cref{sec:appl:wsrmax},
and a more general discussion of this important aspect is provided in \cref{sec:discuss:convergence}.
\end{remark}}

\subsection{Properties of $\mathcal D$ and Implementation of the Feasibility Check} \label{sec:feasibility}
To implement the \cgls{brb} method as described in \cref{alg:bb}, it is necessary to have means to perform the feasibility check in \ref{alg:bb:update} (and \ref{alg:bb:prune}).
Let us first discuss cases in which this can be easily done. Afterwards, we comment on workarounds that can be used if no conclusive feasibility check is available.

A conclusive feasibility test based solely on the properties of \cgls{mm} functions is not possible.
Consider a feasible set
\begin{equation} \label{eq:mmpfeasibleset}
	\mathcal D = \{\vec x \,|\, G_i(\vec x, \vec x) \le 0,\ i = 1, \dots, m\}
\end{equation}
where $G_i$ satisfies \cref{mmp:prop1,mmp:prop2}. These properties lead to the following sufficient conditions for (in-)feasibility of $\mathcal M$.
\begin{proposition}
	Let $\mathcal M = [\vec r, \vec s]$ and $\mathcal D$ as in \cref{eq:mmpfeasibleset}. Then,
	\begin{subequations}
		\label{eq:feas:mmp}
	\begin{align}
		\forall i \in \{1, \ldots, m\} : G_i(\vec s, \vec r) \le 0 &\Rightarrow \mathcal M\cap\mathcal D =\! \mathcal M \!\neq \emptyset \label{eq:feas:mmp:1} \\
		\exists i \in \{1, \ldots, m\} : G_i(\vec r, \vec s) > 0 &\Rightarrow \mathcal M\cap\mathcal D = \emptyset. \label{eq:feas:mmp:2}
	\end{align}
	\end{subequations}
\end{proposition}
\begin{IEEEproof}
	From \cref{mmp:prop1,mmp:prop2}, $G_i(\vec x, \vec x) \le G_i(\vec s, \vec r)$ and $G_i(\vec x, \vec x) \ge G_i(\vec r, \vec s)$ for all $\vec x\in\mathcal M$. Thus, if $G_i(\vec s, \vec r) \le 0$ for all $i = 1, \ldots, l$, then $G_i(\vec x, \vec x) \le 0$ for all $i$ and $\vec x \in \mathcal M$. Hence, \cref{eq:feas:mmp:1}. Similarly, if $G_i(\vec r, \vec s) > 0$ for some $i = 1, \ldots, l$, then also $G_i(\vec x, \vec x) > 0$ for this $i$ and all $\vec x\in\mathcal M$. Thus, $\vec x\notin\mathcal D$ and \cref{eq:feas:mmp:2} holds.
\end{IEEEproof}
In general, there exist boxes for which neither \eqref{eq:feas:mmp:1} nor \eqref{eq:feas:mmp:2} holds,
so that it remains open whether $\mathcal M$ contains a feasible point.
However, we could consider the special case where
\begin{equation} \label{mmp:feas:prop}
	G_i\bigg(\sum_{j\in\mathcal I} x_j \vec e_j, \sum_{k\in\mathcal I^c} y_k \vec e_k \bigg) = G_i(\vec x, \vec y),~~\forall \vec x, \vec y\in\mathds R^n
\end{equation}
for some index set $\mathcal I \subseteq\{1, \dots, n\}$ and all $i = 1, \dots, m$ where $\mathcal I^c = \{1, \dots, n\}\setminus\mathcal I$.
That is, each function $g_i(\vec x) = G_i(\vec x, \vec x)$ is nondecreasing in the variables $x_j$, $j\in\mathcal I$, and nonincreasing in the remaining variables $x_k$, $k\in\mathcal I^c$.
In this case, the following proposition is a simple feasibility test based on \cgls{mm} properties.
\begin{proposition}\label{feas:mmp}
	Let $\mathcal M = [\vec r, \vec s]$ and $\mathcal D$ be defined as in \cref{eq:mmpfeasibleset} by \cgls{mm} functions $G_i(\vec x, \vec y)$ satisfying \cref{mmp:prop12,mmp:feas:prop}. Then, $\mathcal M\cap\mathcal D\neq\emptyset$ if and only if $G_i(\vec r, \vec s) \le 0$ for all $i = 1, \dots, m$. In that case, $\sum_{j\in\mathcal I} r_j \vec e_j + \sum_{k\in\mathcal I^c} s_k \vec e_k \in \mathcal M\cap\mathcal D$ with $\mathcal I$ and $\mathcal I^c$ as in \cref{mmp:feas:prop}.
\end{proposition}
\begin{IEEEproof}
	Let $\vec\xi = (r_1, \dots, r_\kappa, s_{\kappa+1}, \dots, s_K)^\Tr$.
	Then, for all $i$ and due to \cref{mmp:feas:prop},
	$G_i(\vec\xi, \vec\xi) = G_i(\vec r, \vec s)$. Thus, if $G_i(\vec r, \vec s) \le 0$, then $\vec\xi\in\mathcal D$. Since, trivially, $\vec\xi\in\mathcal M$, $\vec\xi\in\mathcal D\cap\mathcal M\neq\emptyset$. Finally, from \cref{eq:feas:mmp:2} follows $\mathcal M\cap\mathcal D\neq\emptyset \Rightarrow G_i(\vec r, \vec s) \le 0$.
\end{IEEEproof}

\begin{corollary}\label{feas:normal}
	Let $\mathcal M = [\vec r, \vec s]$ and $\mathcal D$ be a normal set, i.e.,
	\begin{equation}
		\mathcal D = \{\vec x \,|\, g_i(\vec x) \le 0,\ i = 1, \dots, m\}
	\end{equation}
	with $g_i$ being nondecreasing functions.
	Then, $\mathcal D\cap\mathcal M \neq \emptyset$ if and only if $g_i(\vec r) \le 0$ for all $i = 1, \dots, m$.
\end{corollary}

\begin{corollary}\label{feas:conormal}
	Let $\mathcal M = [\vec r, \vec s]$ and $\mathcal D$ be a conormal set, i.e.,
	\begin{equation}
		\mathcal D = \{\vec x \,|\, h_i(\vec x) \ge 0,\ i = 1, \dots, m\}
	\end{equation}
	with $h_i$ being nondecreasing functions.
	Then, $\mathcal D\cap\mathcal M \neq \emptyset$ if and only if $h_i(\vec r) \ge 0$ for all $i = 1, \dots, m$.
\end{corollary}

\cref{feas:mmp,feas:normal,feas:conormal} cover a wide range of feasible sets.
However, none of these properties is necessary as long as we have other means to perform a feasibility check.
For instance, consider the case where we can express $\mathcal D$ by
\begin{equation} \label{eq:Dfunctional}
	g_i(\vec x) \le 0,\enskip i = 1, \ldots, m, \quad
	h_j(\vec x) = 0,\enskip j = 1, \ldots, l
\end{equation}
where $g_i$ are convex functions and $h_j$ are affine functions. In this case, $\mathcal D$ is a closed convex set and the feasibility check can be solved with polynomial complexity by standard tools from convex optimization \cite{Nesterov1994,BoVa09}.
In particular, \eqref{eq:Dfunctional} includes polyhedral sets where $g_i$ are affine functions.

Let us now discuss workarounds for cases where a feasibility test as described above is not available, but the constraints can be written as \cgls{mm} functions as in \eqref{eq:mmpfeasibleset}.
The first possible workaround is to alter \cref{alg:bb} such that, in \ref{alg:bb:update}, a feasible point is only required if available, and, in \ref{alg:bb:prune}, boxes are only pruned if \cref{eq:feas:mmp:2} is met. Due to this modification, we can use \cref{feas:mmp} instead of a fully conclusive feasibility test.
Then, by a similar argument as in \cite[Prop.~7.4]{Tuy2005} and according to the proof of \cref{thm:bbconv}, there exists an infinite decreasing sequence of sets $\{\mathcal M_{k_q}\}_q$ such that $G_i(\vec r^{k_q}, \vec s^{k_q}) \le 0$, for all $i$ and $q = 1, 2, \ldots$. Since $\diam\mathcal M_{k_q} \rightarrow 0$, $\vec r^{k_q}$ and $\vec s^{k_q}$ approach a common limit point $\vec x$. Due to the continuity of $G_i$, this point satisfies $G_i(\vec x, \vec x) \le 0$ for all $i$. Altering \cref{thm:bbconv}'s proof accordingly, it can be shown that the modified algorithm is infinite and, whenever it generates an infinite sequence $\{ \vec r^k \}$, every accumulation point of this sequence is a global optimum. \chg{Please refer to \cite[Sec.~6.3.1]{diss} for more details.}

In practice, an infinite algorithm often converges in finite time (see the numerical example in \cref{sec:aloha}), but there are no theoretical guarantees for this, and for some problem instances, the resulting algorithm can have very slow convergence.

Another widely accepted workaround is to accept an $\eta$-optimal point that is approximately feasible as solution, i.e., a point $\bar{\vec x}$ satisfying $f(\bar{\vec x}) \ge f(\vec x) - \eta$ for all $\vec x\in\mathcal D$ and $G_i(\bar{\vec x}, \bar{\vec x}) \le \varepsilon$ for all $i = 1, \dots, m$ and some small $\varepsilon > 0$.
Such a point is called \emph{$(\varepsilon, \eta)$-approximate optimal solution}.

This second method restores finite convergence, but gives rise to numerical problems. If $\varepsilon$ is not chosen sufficiently small, the $(\varepsilon, \eta)$-approximate optimal solution might be far from the true optimum. The issue is that it is usually unclear how small is ``sufficient'' to guarantee a good approximate solution \cite[Sec.~7.5]{Tuy2016}. Even worse, if the true optimum is an isolated point,\footnote{A feasible point is called isolated if it is at the center of a ball containing no other feasible points. Please refer to \cite{Matthiesen2018a} for a numerical example showing the existence of isolated feasible points in a radio resource allocation problem.} any change in the tolerances $\varepsilon, \eta$ can lead to drastic changes in the $(\varepsilon, \eta)$-approximate optimal solution \cite[Sec.~4]{Tuy2009}.
We thus generally do not recommend the $(\varepsilon, \eta)$-approximate approach.

A more suitable method for optimization problems with such ``hard'' feasible sets is the \cgls{sit} scheme from \cite{Tuy2009}, which algorithmically excludes all isolated feasible points and provides an elegant solution to the feasibility check issues. An optimization framework based on this scheme is published in \cite{Matthiesen2018a} along with source code, and could be combined with the \cgls{mmp} concept.
Besides its numerical stability, this scheme also improves efficiency for problems that are only nonconvex due to some of their variables \cite{Matthiesen2018a}.

We stress the fact that none of the above workarounds is required if a fully conclusive feasibility test can be implemented (preferably with low computational complexity),
so that the unmodified algorithm as stated in \cref{alg:bb} can be used.

\subsection{The Reduction Procedure in \ref{alg:bb:reduction} of \cref{alg:bb}}
\label{sec:appl:red}
In \ref{alg:bb:reduction} of \cref{alg:bb}, each box $\mathcal M\in\mathscr P_k$ is replaced by a smaller box $\mathcal M'$ that still contains all feasible points that might improve the current best known solution. This step speeds up the convergence since smaller boxes result in tighter bounds. However, it also increases the computation time per iteration and, thus, slows down the algorithm. Ultimately, it depends on the problem at hand, especially the structure of the feasible set $\mathcal D$, and the implementation of the reduction procedure whether \ref{alg:bb:reduction} speeds up \cref{alg:bb} or not. Hence, an important observation is that \ref{alg:bb:reduction} is entirely optional since choosing $\mathcal M' = \mathcal M$ satisfies the above condition. Moreover, note that \cref{eq:alg:bb:reduction} is also satisfied if $\mathcal M'$ satisfies $(\mathcal M\setminus\mathcal M') \cap \mathcal D = \emptyset$.

For $\mathcal D$ convex (or even linear), we refer the reader to the vast literature on convex (or linear) optimization regarding possible implementations of the reduction. Here, we just mention the most straightforward approach, namely to solve the convex (linear) optimization problems
\begin{align}
	r_i' &= \min_{\vec x\in\mathcal M \cap\mathcal D} x_i&
	s_i' &= \max_{\vec x\in\mathcal M \cap\mathcal D} x_i
\end{align}
for all $i = 1, \ldots, n$, and let $\mathcal M' = [\vec r', \vec s']$.

For a feasible set defined by \cgls{mm} constraints as in \cref{eq:mmpfeasibleset}, the reduction can be carried out in a similar fashion as for \cgls{dm} programming problems \cite[Sec.~11.2.1]{Tuy2016}. Let $\mathcal M = [\vec r, \vec s]$ and observe from \cref{eq:feas:mmp:2} that if, for some $i = 1, \ldots, m$, $G_i(\vec r, \vec s) > 0$, then $\mathcal M \cap \mathcal D = \emptyset$ and $\mathcal M' = \emptyset$. Moreover, if $F(\vec s, \vec r) \le \gamma_k$, then $\{\vec x\in\mathcal M \,|\, F(\vec x, \vec x) > \gamma_k \} = \emptyset$ and $\mathcal M' = \emptyset$ satisfies \cref{eq:alg:bb:reduction}. Otherwise, i.e., if $G_i(\vec r, \vec s) \le 0$ for all $i$ and $F(\vec s, \vec r) \ge \gamma_k$, let $\mathcal M' = [\vec r', \vec s']$ with
{\setlength{\belowdisplayskip}{3pt}\begin{align}
\label{eq:reduction:rs}
	\vec r' &= \vec s - \sum_{i=1}^n \alpha_i (s_i - r_i) \vec e_i,~ & 
	\vec s' &= \vec r' + \sum_{i=1}^n \beta_i (s_i - r_i') \vec e_i 
\end{align}}%
and, for all $i = 1, \dots, n$,
{\small\begin{subequations}
\label{eq:reduction:alphabeta}
\begin{flalign}
	&\alpha_i = \sup \mathrlap{\begin{multlined}[t]\Big\{
			\alpha \in [0, 1] \,\Big|\,
		F(\vec s - \alpha (s_i - r_i) \vec e_i, \vec r) > \gamma_k, \\
		G_j(\vec r, \vec s - \alpha (s_i - r_i) \vec e_i) \le 0,\ j = 1, \ldots, m \Big\}
	\end{multlined}}& \label{eq:reduction:alpha} \\
	&\beta_i = \sup \mathrlap{\begin{multlined}[t]\Big\{
			\beta \in [0, 1] \,\Big|\,
		F(\vec s, \vec r' + \beta (s_i - r_i') \vec e_i) > \gamma_k, \\
		G_j(\vec r' + \beta (s_i - r_i') \vec e_i, \vec s) \le 0,\ j = 1, \ldots, m \Big\}.
	\end{multlined}}& \label{eq:reduction:beta}
\end{flalign}
\end{subequations}}%
The proof that \cref{eq:alg:bb:reduction} holds for this reduction procedure is an extension of \cite[Lem.~11.1]{Tuy2016} and \chg{can be found in \cite[Sec.~6.4]{diss}.}
Equations~\cref{eq:reduction:alpha,eq:reduction:beta} can be implemented efficiently by a low precision bisection. It is important though that the obtained solutions are greater (or equal) than the true $\alpha_i$, $\beta_i$. Otherwise, feasible solutions might be lost.

\section{Application Examples}
\label{sec:appl}

\newcommand{\gaina}{\alpha}
\newcommand{\gainb}{\beta}
\newcommand{\noisevar}{\sigma^2}
\newcommand{\eemu}{\mu}
\newcommand{\eepsi}{\Psi}
\newcommand{\eebw}{B}
\newcommand{\rweight}{w}
\newcommand{\tsweight}{\tau}
\newcommand{\misobf}{b}
\newcommand{\misoa}{\alpha}
\newcommand{\misob}{\beta}
\newcommand{\misoparam}{\zeta}
\newcommand{\tsmu}{w}
\newcommand{\tslambda}{\nu}
\newcommand{\utility}{U}
\newcommand{\alohaP}{\theta}
\newcommand{\alohaR}{c}
\newcommand{\alohaY}{\upsilon}
\newcommand{\alohaYset}{\mathcal{Y}}
\newcommand{\rate}{r}
\newcommand{\mmrate}{R}
\newcommand{\ratevar}{\rho}
\newcommand{\Rate}{R}
\newcommand{\rmink}[1][k]{R_{\mt{min},#1}}
\newcommand{\sinr}{\gamma}
\newcommand{\sinrreg}{\mathcal{G}}
\newcommand{\sinrmink}[1][k]{\gamma_{\mt{min},#1}}
\newcommand{\ratereg}{\mathcal{R}}

To demonstrate the usefulness and exceptional performance of the proposed \cgls{mmp} approach, we consider examples of various applications in the area of signal processing for communications.
Where available, existing globally optimal approaches are discussed and compared to the proposed framework.
\chg{Run} time comparisons show tremendous gains over the state-of-the-art solutions.
For the example considered in \cref{sec:appl:wsrmax}, we even provide an analytical justification why the proposed method outperforms previous \cgls{dm} formulations.

The complete source code is available on GitHub \cite{github}. All reported performance results were obtained on Intel Haswell nodes with Xeon E5-2680 v3 CPUs running at \SI{2.50}{\giga\Hz}.

The presented applications are only meant as examples:
we are aware of further optimization problems for which the \cgls{mmp} framework can be useful, and we are convinced that further applications can also be identified in other research areas.

\subsection{Weighted Sum Rates in the $K$-User Interference Channel}
\label{sec:appl:wsrmax}
As a first application example, we consider weighted sum rate maximization in a $K$-user \cgls{ic}
under the assumption that the input signals are proper Gaussian and that interference is treated as noise.
Letting $\gaina_k$ denote the gains of the intended channels, and $\gainb_{k\notk}$ with $\notk\neq k$ the gains of the unintended channels,
we can write the achievable rates as 
\begin{equation}
\label{eq:tinrate}
\rate_k = \log_2 \left( 1+ \frac{\gaina_k \pow_k}{\noisevar + \sum_{\notk = 1}^K \gainb_{k\notk} \pow_\notk } \right)
\end{equation}
where $\noisevar$ is the noise variance and $\pow_k$ is the transmit power of user $k$.
Due to the possibility of modeling self-interference or hardware impairments by choosing $\gainb_{kk} \neq 0$,
this formulation is more general than in some of the previous works mentioned below.
Note that there are several other system models for which the rate expressions can be brought to a form equivalent to \eqref{eq:tinrate},
e.g., certain massive \cgls{mimo}, cellular, and relay-aided scenarios \cite{Zappone2016,wcnc18}.

The weighted sum rate maximization problem with minimum rate constraints is
\begin{equation}
\label{eq:appl:wsrmax:prob}
\max_{\vec 0 \le \vec p \le \vec P}~ \sum_{k=1}^K \rweight_k  \rate_k
\quad\st\quad \rate_k \geq \rmink,\ k = 1, \dots, K.
\end{equation}
For this problem, various approaches can be found in the literature.
In the MAPEL framework \cite{Qian2009,Zhang2012}, the problem is parametrized in terms of \cglspl{sinr} as
\begin{equation}
\label{eq:appl:wsrmax:prob_mapel}
\max_{(\sinr_k)_\allk \in \sinrreg }~ \sum_{k=1}^K \rweight_k  \log_2(1+\sinr_k)
\quad\st\quad \sinr_k \geq \sinrmink,~\allk
\end{equation}
where the set of possible \cgls{sinr} combinations $\sinrreg$ is approximated from the outside by means of the \cgls{pa} \cite{Tuy2000} until the global optimal solution is found.
Instead, the authors of \cite{Liu2012} formulate the problem as
\begin{equation}
\label{eq:appl:wsrmax:prob_ratespace}
\max_{\vec\ratevar \in \ratereg }~ \sum_{k=1}^K \rweight_k  \ratevar_k
\quad\st\quad \ratevar_k \geq \rmink,\ k = 1, \dots, K
\end{equation}
where $\ratereg$ is the achievable rate region \chg{defined by} \cref{eq:tinrate} and the power constraints. The rate region is then approximated
by the \cgls{pa}. This special case of the framework \chg{in} \cite{Br12,UtBr12} is termed as ``Ratespace \cgls{pa}'' in the numerical results below.
A disadvantage of both methods is that in every iteration an inner problem with considerable computational complexity has to be solved to project points from outside the feasible set onto its boundary.

Another approach to apply the monotonic optimization framework \cite{Tuy2000} is
to rewrite the rates as \cgls{dm} functions
\begin{equation}
\label{eq:tinrate_dm}
\rate_k = \log_2 \Bigg( \gaina_k \pow_k + \noisevar + \sum_{\notk = 1}^K \gainb_{k\notk} \pow_\notk \Bigg) -
\log_2 \Bigg( \noisevar + \sum_{\notk = 1}^K \gainb_{k\notk} \pow_\notk \Bigg).
\end{equation}
This problem can either be solved via the \cgls{pa} by introducing an auxiliary variable \cite{Matthiesen2017,Tuy2000} (termed as ``\cgls{pa}'') or directly via the \chg{\cgls{bb}} method for \cgls{dm} problems \cite{Tuy2005} (``\chg{BB} DM'').\footnote{\chg{The \cgls{bb} algorithm in \cite{Tuy2005} includes a reduction step similar to the one described in \cref{sec:appl:red}. However, for the problem under consideration, omitting the reduction step leads to faster performance.}}
Among the state-of-the-art, this \chg{\cgls{bb}} approach is most closely related to the proposed \cgls{mmp} framework. Indeed, finding an \cgls{mmp} representation of \cref{eq:tinrate_dm} is straightforward \chg{as explained in \cref{sec:prob}: all powers in the first $\log$-term of \cref{eq:tinrate_dm} are replaced} by nondecreasing variables $x_i$, and all powers in the second $\log$-term \chg{are replaced} by nonincreasing variables $y_i$. However, we will show below in \cref{eq:appl:wsrmax:compareMMPDM} that this leads to looser bounds and, thus, slower average convergence speed than the new \cgls{mmp} representation proposed below.

By calculating the partial derivatives of $\rate_k$ in \eqref{eq:tinrate}, it is easy to verify that $\rate_k$
is nondecreasing in $\pow_k$ (regardless of the value of $\pow_\notk$) and nonincreasing in $\pow_\notk$ for $\notk\neq k$ (regardless of \chg{the} value of $\pow_k$).
Thus, the \cgls{mm} function
\begin{equation}
\label{eq:tinrate_mm}
\mmrate_k(\brbx,\brby) = \log_2 \left( 1+ \frac{\gaina_k \brbxk[k]}{\noisevar + \gainb_{kk}\brbxk[k] + \sum_{\notk \neq k} \gainb_{k\notk} \brbyk[\notk] } \right)
\end{equation}
is an \cgls{mmp} representation of \cref{eq:tinrate}.
Using \cref{eq:mmrules:sum,eq:mmrules:incdec}, an \cgls{mmp} representation of the objective of \eqref{eq:appl:wsrmax:prob} is obtained as
\begin{equation} \label{eq:appl:wsrmax:mmpobjective}
\brbF(\brbx,\brby) = \sum_{k=1}^K \rweight_k \mmrate_k(\brbx,\brby).
\end{equation}
An \cgls{mmp} formulation of the feasible set is given by \cref{eq:mmpfeasibleset} with
\begin{equation}
	\brbG_k(\brbx,\brby) = \rmink - \mmrate_k(\brby,\brbx),\ k = 1, \ldots, K.
\end{equation}

The average convergence speed of \cgls{bb} methods depends strongly on the quality of the bounds, i.e., tighter bounds lead, in general, to faster \chg{convergence \cite{Tuy2005}}. Consider a single rate $\rate_k$ and the bounds obtained by \cref{eq:tinrate_dm,eq:tinrate_mm} evaluated for a box $\brbbox{\brbx}{\brby}$. The difference between \cref{eq:tinrate_mm} and \cref{eq:tinrate_dm} is
\begin{subequations}
\label{eq:appl:wsrmax:compareMMPDM}
\begin{align}
&\label{eq:appl:wsrmax:compareMMPDM:approach}
\mmrate_k(\brbx,\brby) \,-\\\nonumber
& \log_2 \Bigg( \gaina_k \brbxk[k] + \noisevar + \sum_{\notk = 1}^K \gainb_{k\notk} \brbxk[\notk] \Bigg) -
\log_2 \Bigg( \noisevar + \sum_{\notk = 1}^K \gainb_{k\notk} \brbyk[\notk] \Bigg)
\\=\,
&\log_2 \left( \frac
{ \gaina_k \brbxk[k] + \noisevar + \gainb_{kk}\brbxk[k] + \sum_{\notk\neq k}^K \gainb_{k\notk} \brbyk[\notk]}
{ \gaina_k \brbxk[k] + \noisevar + \gainb_{kk}\brbxk[k] + \sum_{\notk\neq k}^K \gainb_{k\notk} \brbxk[\notk]}
 \right)
\,+\nonumber\\&
\label{eq:appl:wsrmax:compareMMPDM:result}
\log_2 \left( \frac
{\noisevar + \gainb_{kk}\brbyk[k] + \sum_{\notk\neq k}^K \gainb_{k\notk} \brbyk[\notk] }
{\noisevar + \gainb_{kk}\brbxk[k]  + \sum_{\notk\neq k} \gainb_{k\notk} \brbyk[\notk]}
\right)
\leq0
\end{align}
\end{subequations}
where we have exploited that  
the numerator is \chg{greater or equal} than the denominator in both fractions due to
$\brby \geq \brbx$.
This shows that the \cgls{mmp} bound is always tighter than the \cgls{dm} bound.\footnote{For problems without self-interference,
the above comparison simplifies as the second logarithm in \eqref{eq:appl:wsrmax:compareMMPDM:result} vanishes,
but the conclusion remains the same.}

\begin{figure}
	\centering
	\tikzpicturedependsonfile{tin_iterations.dat}
	\tikzsetnextfilename{tinWsrIterations}
	\begin{tikzpicture}
		\begin{semilogyaxis} [
				thick,
				xlabel={Users},
				ylabel={Iterations},
				ylabel near ticks,
				xlabel near ticks,
				grid,
				xminorgrids,
				yminorgrids,
				xmax = 18,
				xmin = 2,
				minor x tick num = 2,
				xtick distance = 3,
				ytick = {1e2, 1e4, 1e6, 1e8},
				minor ytick = {1e3, 1e5, 1e7},
				mark repeat = 1,
				legend pos=south east,
				legend style={font=\small},
				legend cell align=left,
				cycle list name=default,
				width = \axisdefaultwidth,
				height = .63*\axisdefaultheight,
			]

			\pgfplotstableread[col sep=comma]{tin_iterations.dat}\tbl

			\addplot table[y=mmp] {\tbl};
			\addlegendentry{MMP};

			\addplot table[y=mmpFifo] {\tbl};
			\addlegendentry{MMP oldest-first};

			\addplot table[y=dm] {\tbl};
			\addlegendentry{BRB DM};
		\end{semilogyaxis}
	\end{tikzpicture}
	\vspace{-2ex}
	\caption{Average number of iterations in \cref{alg:bb} to solve \cref{eq:appl:wsrmax:prob} with bounds obtained by \cref{eq:tinrate_mm} with best- and oldest-first selection, respectively, and \cref{eq:tinrate_dm} with best-first selection. Results are averaged over 100 \cgls{iid} channel realizations.}\vspace*{-2mm}
	\label{fig:tiniterations}
\end{figure}
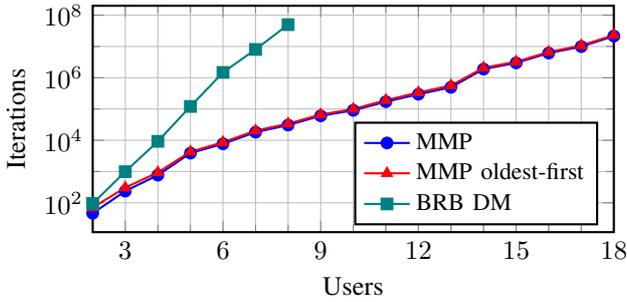

The results in \cref{fig:tiniterations} illustrate this theoretical intuition numerically. The plot displays the average number of iterations required by \cref{alg:bb} to solve \cref{eq:appl:wsrmax:prob} versus the number of variables. Each data point is averaged over $100$ \cgls{iid} channel realizations with $\alpha_i = \abs{\alpha_i'}^2$ and $\beta_{ij} = \abs{\beta_{ij}'}^2$ where $\alpha_i', \beta_{ij}' \sim \mathcal{CN}(0, 1)$ for all $i$ and $j\neq i$. Further, $\eta = 0.01$, $\sigma^2 = 0.01$, $P_k = 1$, $w_k = 1$, $\beta_{kk} = 0$, $\rmink=0$ for all $k$, and no reduction is used. It can be observed that \cgls{mmp} with best-first selection \eqref{eq:selection:argmax} and \cgls{mm} function \cref{eq:tinrate_mm}, labeled as ``\cgls{mmp},'' requires three orders of magnitude less iterations than the same algorithm with \cgls{mm} function obtained from \cref{eq:tinrate_dm}, named ``\chg{BB} DM,'' to solve \cref{eq:appl:wsrmax:prob} with $K = 8$ variables. From a practical perspective, this means that the \cgls{mmp} framework is able to solve \cref{eq:appl:wsrmax:prob} with $18$ variables in the same time that state-of-the-art monotonic programming requires for $8$ variables. A further observation from \cref{fig:tiniterations} is that the oldest-first selection rule \eqref{eq:selection:oldest}, labeled as ``\cgls{mmp} oldest-first,'' requires only slightly more iterations to converge than the best-first rule. The benefits of the oldest-first selection will be further evaluated below.

\begin{figure}
	\centering
	\tikzpicturedependsonfile{tin_runtime.dat}
	\tikzsetnextfilename{tinWsrBenchmark}
	\begin{tikzpicture}
		\begin{semilogyaxis} [
				thick,
				xlabel={Users},
				ylabel={Run Time [s]},
				ylabel near ticks,
				xlabel near ticks,
				grid,
				xminorgrids,
				xmax = 18,
				xmin = 2,
				ymin = 10^-5,
				ymax = 10^5,
				minor x tick num = 2,
				xtick distance = 3,
				ytick distance = 10^1,
				mark repeat = 1,
				legend pos=north east,
				legend style={font=\small},
				legend columns = 2,
				legend cell align=left,
				cycle list name=default,
			]

			\pgfplotstableread[col sep=comma]{tin_runtime.dat}\tbl

			\addplot table[y=mmp] {\tbl};
			\addlegendentry{MMP};

			\addplot table[y=mmpFifo] {\tbl};
			\addlegendentry{MMP oldest-first};

			\addplot table[y=dm] {\tbl};
			\addlegendentry{BB DM};

			\addplot table[y=pa] {\tbl};
			\addlegendentry{PA};

			\addplot table[y=mapel] {\tbl};
			\addlegendentry{MAPEL};

			\addplot table[y=ratespace] {\tbl};
			\addlegendentry{Ratespace PA};
		\end{semilogyaxis}
	\end{tikzpicture}
	\vspace{-2ex}
\caption{Average run time required to solve \cref{eq:appl:wsrmax:prob} with different algorithms. Results are averaged over 100 different \cgls{iid} channel realizations.}
	\label{fig:tinruntime}
\end{figure}
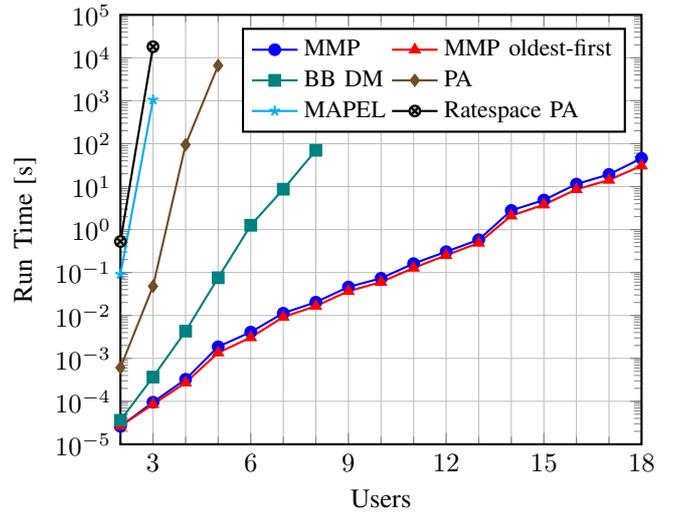

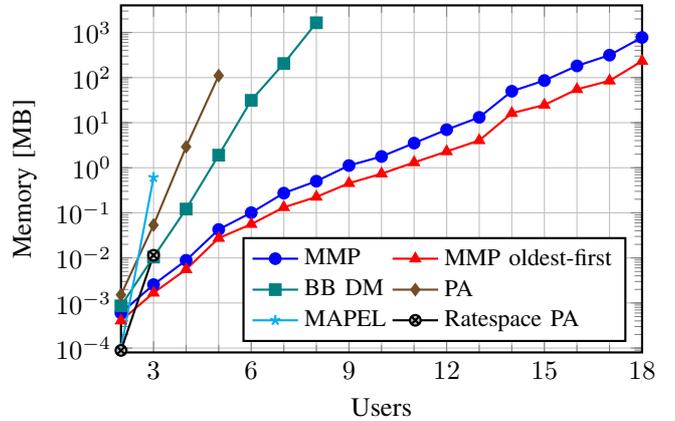
\begin{figure}
	\centering
	\tikzpicturedependsonfile{tin_memory.dat}
	\tikzsetnextfilename{tinWsrBenchmarkMem}
	\begin{tikzpicture}
		\begin{semilogyaxis} [
				thick,
				xlabel={Users},
				ylabel={Memory [MB]},
				ylabel near ticks,
				xlabel near ticks,
				grid,
				xminorgrids,
				xmax = 18,
				xmin = 2,
				ymax = 4e3,
				ymin = 8e-5,
				minor x tick num = 2,
				xtick distance = 3,
				ytick distance = 10^1,
				mark repeat = 1,
				legend pos=south east,
				legend cell align=left,
				legend style={font=\small},
				legend columns = 2,
				cycle list name=default,
				width = \axisdefaultwidth,
				height = .85*\axisdefaultheight,
			]

			\pgfplotstableread[col sep=comma]{tin_memory.dat}\tbl

			\addplot table[y expr=\thisrow{mmp}/1e6] {\tbl};
			\addlegendentry{MMP};

			\addplot table[y expr=\thisrow{mmpFifo}/1e6] {\tbl};
			\addlegendentry{MMP oldest-first};

			\addplot table[y expr=\thisrow{dm}/1e6] {\tbl};
			\addlegendentry{BB DM};

			\addplot table[y expr=\thisrow{pa}/1e6] {\tbl};
			\addlegendentry{PA};

			\addplot table[y expr=\thisrow{mapel}/1e6] {\tbl};
			\addlegendentry{MAPEL};

			\addplot table[y expr=\thisrow{ratespace}/1e6] {\tbl};
			\addlegendentry{Ratespace PA};
		\end{semilogyaxis}
	\end{tikzpicture}
	\vspace{-2ex}
	\caption{Average memory consumption of different algorithms to solve \cref{eq:appl:wsrmax:prob}. Results are averaged over 100 different \cgls{iid} channel realizations.}\vspace*{-2mm}
	\label{fig:tinmemory}
\end{figure}

A comparison based on iterations works well for algorithms with similar computational complexity per iteration. However, when \chg{evaluating} algorithms as different as \chg{\cgls{bb} algorithms and the \cgls{pa}}, comparing the number of iterations is meaningless: the \cgls{pa} typically requires much less iterations but each iteration takes much longer than in a \cgls{bb} algorithm. Thus, we resort to \chg{measuring} the average run time of the algorithms in the C++ implementation available at \cite{github}. We have taken great care to implement the state-of-the-art algorithms with the same rigor and amount of code optimization as the proposed method to make this benchmark as fair as possible. The average run time and memory consumption of all discussed approaches is displayed in \cref{fig:tinruntime,fig:tinmemory}, respectively. The same parameters as in the computation of \cref{fig:tiniterations} were used.

First, observe that all algorithms scale both in run time and memory consumption exponentially in the number of variables. Since problem~\cref{eq:appl:wsrmax:prob} is NP hard \cite[Thm.~1]{Luo2008}, better asymptotic complexity is most likely not achievable. However, it is obvious that the computational complexity still may have very different slope and some algorithms are significantly more efficient than others. The 
proposed \cgls{mmp} framework solves problem~\cref{eq:appl:wsrmax:prob} in considerably less time and memory requirements than all other state-of-the-art methods. The \cgls{pa} based methods all consume more memory than the \cgls{bb} based methods starting from three optimization variables. In terms of run time, they are already outperformed by at least 1.5 orders of magnitude for 2 variables and soon reach our run time limit of \SI{8}{\hour}. For the \cgls{bb} methods, the observations from \cref{fig:tiniterations} continue to hold in \cref{fig:tinruntime}. From the memory consumption, it can be observed that good bounds are not only critical for fast convergence but also for memory efficiency. In this example, the \cgls{mmp} method was able to solve problems more than twice the size of the \cgls{dm} \cgls{bb} method within a memory limit of \SI{2.5}{\giga\byte}.

Finally, observe that the best-first approach consumes considerably more memory than the oldest-first rule, e.g., 3.4$\times$ or \SI{546}{\mega\byte} more at 18~variables. Further, observe that while requiring less iterations, the best-first rule has longer run times than the oldest-first rule. This can be explained from \cref{fig:tinmemory} since the memory consumption is directly proportional to the number of boxes in $\mathcal R_k$. The best-first rule is the mathematical description of a priority queue. While accessing the top-element in a priority queue has complexity $O(1)$, insertion has worst-case complexity $O(\log n)$ \cite[pp.~148--152]{Knuth1997vol3}, where $n$ is the number of elements in the data structure.
Compared to the other operations during each iteration of the algorithm, which have polynomial complexity in the number of variables, $O(\log n)$ is extremely small except when the size of the queue is very large.
Instead, the implementation of the oldest-first rule is a queue, i.e., a \gls{fifo} list. 
Here, the insertion, deletion, and access to the front element all require constant time $O(1)$ and do not grow with the number of stored elements.

\subsection{Energy Efficiency Optimization} \label{sec:appl:gee}
The \cgls{gee} is a key performance metric for 5G and beyond networks measuring the network energy efficiency \cite{Isheden2012,Zappone2015,Zappone2016}. 
It is defined as the benefit-cost ratio of the total network throughput in a time interval $T$ and the energy necessary to operate the network during this time:
\begin{equation}
	\mathrm{GEE} = \frac{T B \sum_{k=1}^K r_k}{T \left( \vec\phi^\Tr \vec p + P_c \right)} = \frac{B \sum_{k=1}^K r_k}{\vec\phi^\Tr \vec p + P_c} \quad \left[ \frac{\mathrm{bit}}{\mathrm{J}} \right],
	\label{eq:gee}
\end{equation}
where $r_k$ is the achievable rate of link $k$, $B$ is the bandwidth, $\vec\phi\ge \one$ contains the inverses of the power amplifier efficiencies and $P_c$ is a constant modeling the constant part of the circuit power consumption.

Maximizing the \cgls{gee} for interference networks with \cgls{tin}, i.e., where $r_k$ is as in \cref{eq:tinrate}, results in the nonconvex fractional programming problem \cite{Schaible1993,Zappone2015}
\begin{equation}
	\label{eq:geemax}
	\max_{\vec 0 \le \vec p \le \vec P}\enskip \frac{\sum_{k=1}^K r_k}{\vec\phi^\Tr \vec p + P_c}
\end{equation}
where we have omitted the inessential constant $B$ and minimum rate constraints that are already discussed in \cref{sec:appl:wsrmax}.
As the objective includes the sum rate as a special case for $\vec\phi=\vec 0$ and $P_c = 1$, this problem is also NP-hard due to \cite[Thm.~1]{Luo2008}. As already mentioned below \cref{eq:frac}, the state-of-the-art approach to solve \cref{eq:geemax} is to combine Dinkelbach's Algorithm \cite{dinkelbach1967,Zappone2015} with monotonic programming. This was first proposed in \cite{Matthiesen2015} and subsequently developed into the fractional monotonic programming framework in \cite{Zappone2017}. Dinkelbach's Algorithm solves \cref{eq:geemax} as a sequence 
of auxiliary problems
\begin{equation}
	\label{eq:dinkelbachaux}
	\max_{\vec 0 \le \vec p \le \vec P}\enskip \sum_{k=1}^K r_k - \lambda \vec\phi^\Tr \vec p + P_c
\end{equation}
with non-negative parameter $\lambda$. Problem~\cref{eq:dinkelbachaux} can be solved by monotonic programming much in the same way as discussed in \cref{sec:appl:wsrmax}. While most works use the \cgls{pa} to solve \cref{eq:dinkelbachaux} (e.g., \cite{Matthiesen2015,Zappone2017}), we have already demonstrated above that \chg{\cgls{bb} with \cgls{dm} bounds \cite{Tuy2005}} outperforms the classical \cgls{pa} \cite{Tuy2000}.

The \cgls{mmp} framework even allows to solve \cref{eq:geemax} without the need of Dinkelbach's Algorithm. An \cgls{mmp} representation of \cref{eq:gee} can be obtained similar to \cref{eq:frac}. Specifically, with \cref{eq:tinrate_mm} and the identities in \cref{eq:mmrules:sum,eq:mmrules:prod,eq:mmrules:incdec}, we obtain
\begin{equation}
\label{eq:appl:gee_mm}
\brbF(\brbx, \brby) =
\frac{\eebw \sum_{k=1}^K \mmrate_k(\brbx, \brby)}
{\vec\phi^\Tr \brby + P_c  }
\end{equation}
with $\mmrate_k(\brbx, \brby)$ as in \cref{eq:tinrate_mm}.

\begin{figure}
	\centering
	\tikzpicturedependsonfile{gee_runtime.dat}
	\tikzsetnextfilename{tinGeeBenchmark}
	\begin{tikzpicture}
		\begin{semilogyaxis} [
				thick,
				xlabel={Users},
				ylabel={Run Time [s]},
				ylabel near ticks,
				xlabel near ticks,
				grid,
				xminorgrids,
				xmax = 10,
				xmin = 2,
				minor x tick num = 0,
				ytick = {1e-4, 1e-2, 1, 1e2, 1e4},
				minor ytick = {1e-3,1e-1,1e1,1e3},
				xtick distance = 1,
				mark repeat = 1,
				legend pos=north east,
				legend cell align=left,
				legend style={font=\small},
				cycle list name=default,
				width = \axisdefaultwidth,
				height = .55*\axisdefaultheight,
			]

			\pgfplotstableread[col sep=comma]{gee_runtime.dat}\tbl

			\addplot table[y=gee_mmp] {\tbl};
			\addlegendentry{MMP};

			\addplot table[y=gee_dinkelbach] {\tbl};
			\addlegendentry{Dinkelbach};
		\end{semilogyaxis}
	\end{tikzpicture}
	\vspace{-2ex}
	\caption{Average run time to solve \cref{eq:geemax} with \cgls{mmp} and Dinkelbach's Algorithm where the inner problem is solved by \cref{alg:bb} with \cgls{dm} rate expressions \cref{eq:tinrate_dm}. Results are averaged over 100 different \cgls{iid} channel realizations.}\vspace*{-3mm}
	\label{fig:geeruntime}
\end{figure}
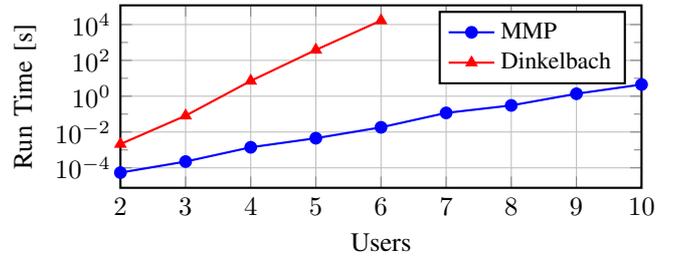

The run time performance of both algorithms is evaluated in \cref{fig:geeruntime} where $\phi_k = 5$, for all $k$, and $P_c = 1$. The remaining parameters were chosen as in \cref{sec:appl:wsrmax}. It can be observed that \cgls{mmp} requires significantly less time to solve \cref{eq:geemax} than the legacy approach employing Dinkelbach's Algorithm. For example, with $K= 6$ variables, \cgls{mmp} is on average almost five orders of magnitude faster than fractional monotonic programming. The memory consumption (not displayed) scales almost identically to the run time with \cgls{mmp} using four orders of magnitude less memory \chg{than Dinkelbach's Algorithm} for six variables. Besides showing much better run time and memory performance, the \cgls{mmp} method also guarantees an $\eta$-optimal solution. By contrast, Dinkelbach's Algorithm does not provide any guarantees on the solution quality
since an inaccuracy of $\eta$ in the inner solver might propagate to larger inaccuracies in the overall results.

Other \cgls{ee} metrics can be maximized with the \cgls{mmp} framework in a similar manner. For example, in interference networks with rate function \cref{eq:tinrate}, the weighted minimum \cgls{ee} (WMEE) has the \chg{objective \cite{Zappone2017}}
\begin{equation}
	\mathrm{WMEE} = \min_{k = 1, \ldots, K} w_k \frac{B r_k}{\phi_k p_k + P_{c,k}}
\end{equation}
with nonnegative weights $w_1, \ldots, w_K$
and \cgls{mmp} representation
\begin{equation}
	F_{\mathrm{WMEE}}(\vec x, \vec y) = \min_{k = 1, \ldots, K} w_k \frac{B R_k(\vec x, \vec y)}{\phi_k y_k + P_{c,k}},
\end{equation}
\chg{and the} weighted sum \cgls{ee} (WSEE) has the \chg{objective \cite{Zappone2017}}
\begin{equation} \label{eq:wsee}
	\mathrm{WSEE} = \sum_{k = 1}^K w_k \frac{B r_k}{\phi_k p_k + P_{c,k}}
\end{equation}
with \cgls{mmp} representation
\begin{equation}
	F_{\mathrm{WSEE}}(\vec x, \vec y) = \sum_{k = 1}^K w_k \frac{B R_k(\vec x, \vec y)}{\phi_k y_k + P_{c,k}}.
\end{equation}

The WMEE can be maximized similarly to the \cgls{gee} with a combination of the Generalized Dinkelbach Algorithm and monotonic programming \cite{Zappone2017}. Instead, optimizing the WSEE with monotonic optimization is much more challenging since neither Dinkelbach's Algorithm nor its generalization are applicable. In \cite{Zappone2017}, it is proposed to transform \cref{eq:wsee} into a single fractional program, i.e.,
\begin{equation}
	\frac{\sum_{k = 1}^K w_k B r_k \prod_{i\neq k} (\phi_i p_i + P_{c,i}) }{\prod_{k=1}^K (\phi_k p_k + P_{c,k})}
\end{equation}
and then apply fractional monotonic programming. While this works in theory, it is shown in \cite{wcnc18} that this approach has very poor convergence. Instead, the \cgls{mmp} framework allows to directly optimize both metrics without cumbersome transformations and without using Dinkelbach's Algorithm or its generalized version.

\subsection{Proportional Fair Rate Optimization with Scheduling}
\label{sec:appl:scheduling}
The weighted sum rate utility in \eqref{eq:appl:wsrmax:prob} can also be replaced by a utility function that accounts for the fairness between users, such as the proportional fair utility
\begin{equation}
\label{eq:PFutility}
\utility \left( \rate_1,\dots,\rate_K   \right) = \sum_{k=1}^K \ln  \rate_k.
\end{equation}
In this context, a common approach (e.g., \cite{Qian2010,Zhang2012,Br12,UtBr12}) is to increase the flexibility in the optimization by scheduling different transmit strategies in multiple time slots and averaging the data rates, i.e.,
\begin{align}
\label{eq:appl:smapel:prob_schedule}
\max_{\substack{(0\leq \pow_k^{(\ms)}\leq P_k)_{\allk\forall\ms}\\\Ms\in\mathbb{N},\tau\geq\zero: \one^\Tr\mb\tau=1}}~
 \utility \left( \bar\rate_1,\dots,\bar\rate_K \right)
 ~\st~ \bar\rate_k\geq \rmink,~\allk
\end{align}
with
\begin{align}
\label{eq:appl:tsrates}
\bar\rate_k &= \sum_{\ms=1}^{\Ms} \tsweight_\ms \rate_k^{(\ms)},
&
\rate_k^{(\ms)} &=
\log_2 \left( 1+ \frac{\gaina_k \pow_k^{(\ms)}}{\noisevar + \sum_{j\neq k}\gainb_{kj} \pow_j^{(\ms)} } \right).
\end{align}
In this application example, we restrict ourselves to the proportional fair utility \eqref{eq:PFutility}
since this problem \chg{was shown to be NP hard \cite{Luo2008} for all $L\ge 3$} even though the utility is concave in the per-user rates.

In \cite{Qian2010,Zhang2012}, an algorithm called S-MAPEL for nondecreasing utility functions was proposed.
The approach is based on the \cgls{pa}
and makes use of the reformulation \eqref{eq:appl:wsrmax:prob_mapel} as well as of the observation that the rates are nondecreasing functions of the time fractions $\tsweight_\ms$.
By arguing that no more than $L=K+1$ strategies are necessary due to the Carath\'eodory theorem,
the approach from \cite{Qian2010,Zhang2012} uses $LK=(K+1)K$ optimization variables in total.\footnote{In fact,
the number of strategies can be reduced to $L=K$ due to an extension to the Carath\'eodory Theorem discussed in \cite{HaRa51},
yielding a total number of $LK=K^2$ variables.}
This leads to a significant computational complexity.
A second disadvantage of this approach is as follows. The optimizer of \eqref{eq:appl:smapel:prob_schedule} is not unique since any re-indexing of the time index $\ms$
leads to an optimal solution as well, but when directly solving \eqref{eq:appl:smapel:prob_schedule} this inherent symmetry is not exploited.
The authors of \cite{Qian2010,Zhang2012} thus proposed an accelerated algorithm called A-S-MAPEL which employs a heuristic (with an additional tolerance parameter $\asmeps$) to exploit the symmetry,
but the resulting strategy is no longer guaranteed to be $\eta$-optimal.

We focus on the following alternative method for concave utility functions from \cite{Br12,UtBr12},
which avoids increasing the number of variables at the cost of having to solve a series of monotonic optimization problems.
To obtain an efficient algorithm, we combine this approach with the \cgls{mmp} framework.

We rewrite problem~\eqref{eq:appl:smapel:prob_schedule} as
\begin{align}
\label{eq:appl:smapel:prob_rewrite}
\max_{\substack{(0\leq \pow_k^{(\ms)}\leq P_k)_{\allk\forall\ms}\\(\ratevar_k\geq \rmink)_{\allk}\\\Ms\in\mathbb{N},\tau\geq\zero: \one^\Tr\mb\tau=1}}~
& \utility \left( \ratevar_1,\dots,\ratevar_K \right)
\quad\st\quad \bar\rate_k\geq\ratevar_k~~\allk
\end{align}
and consider the Lagrangian dual problem
\begin{align}
\label{eq:appl:smapel:prob_dual}
\min_{\mb\mu\geq\zero}
\max_{\substack{(0\leq \pow_k^{(\ms)}\leq P_k)_{\allk\forall\ms}\\(\ratevar_k\geq \rmink)_{\allk}\\\Ms\in\mathbb{N},\tau\geq\zero: \one^\Tr\mb\tau=1}}~
\utility \left( \ratevar_1,\dots,\ratevar_K \right) + \sum_{k=1}^K  \mu_k(\bar\rate_k-\ratevar_k)
\end{align}
where $\bar\rate_k$ depends on the optimization variables via \eqref{eq:appl:tsrates}.
Since averaging the rates can be interpreted as optimizing over the convex hull of the achievable rate region, \eqref{eq:appl:smapel:prob_rewrite} can be rewritten
as a convex program to show that strong duality holds \cite{Br12,UtBr12}, i.e.,
\eqref{eq:appl:smapel:prob_dual} has the same optimal value as \eqref{eq:appl:smapel:prob_rewrite}.

We note that $\mb\pow^{(\ms)}$ can be optimized separately for each $\ms$,
and that these inner problems are all equivalent, i.e., 
\begin{subequations}
\begin{align}
&
\max_{\substack{(0\leq \pow_k^{(\ms)}\leq P_k)_{\allk\forall\ms}\\\Ms\in\mathbb{N},\tau\geq\zero: \one^\Tr\mb\tau=1}} 
\mb\mu^\Tr \sum_{\ms=1}^\Ms \tau_\ms \mb\rate^{(\ms)}
 \\=&~ 
\max_{\Ms\in\mathbb{N},\tau\geq\zero: \one^\Tr\mb\tau=1} 
\underbrace{\sum_{\ms=1}^\Ms \tau_\ms}_{=1}
\max_{(0\leq \pow_k^{(1)}\leq P_k)_{\allk}} 
\mb\mu^\Tr \mb\rate^{(1)}
\end{align}
\end{subequations}
which implies that the choice of $L$ and $\mb\tau$ in the dual problem is arbitrary.
Thus, the dual problem \eqref{eq:appl:smapel:prob_dual} can be rewritten as
%
\newcommand{\dualfA}{u}
\newcommand{\dualfB}{v}
\begin{align}
\label{eq:appl:smapel:prob_dualsplit}
\min_{\mb\mu\geq\zero}~~
\dualfA_{\mb\mu}(\mb\ratevar^\star(\mb\mu))
+
\dualfB_{\mb\mu}(\mb\pow^\star(\mb\mu))
\end{align}
where
\begin{subequations}
\label{eq:appl:smapel:prob_inner}
\begin{align}
\label{eq:appl:smapel:prob_innerA}
\mb\ratevar^\star(\mb\mu)&=\!\!\!\! \argmax_{(\ratevar_k\geq \rmink)_{\allk}} \dualfA_{\mb\mu}(\mb\ratevar),
&
\dualfA_{\mb\mu}(\mb\ratevar) &= \utility ( \mb\ratevar ) - \mb\mu^\Tr \mb\ratevar
\\
\label{eq:appl:smapel:prob_innerB}
\mb\pow^\star(\mb\mu)&= \!\!\!\! \argmax_{(0\leq \pow_k\leq P_k)_{\allk}} \dualfB_{\mb\mu}(\mb\pow),
&
\dualfB_{\mb\mu}(\mb\pow) &= \mb\mu^\Tr \mb\rate.
\end{align}
\end{subequations}

In total, we have to solve three optimization problems in \eqref{eq:appl:smapel:prob_dualsplit}.
The outer minimization is a convex problem in the dual variables $\mb\mu$ and \chg{can be} solved by the cutting plane method \cite{BaShSh06,Ke60}
which successively refines outer approximations
\begin{subequations}
\label{eq:appl:smapel:dual_CP}
\begin{align}
&\min_{\substack{\mb\mu\geq\zero,z\in\mathbb{R}}} ~~ z 
\\&\st~~
z\geq 
\dualfA_{\mb\mu}(\mb\ratevar^{(\ms)})
+
\dualfB_{\mb\mu}(\mb\pow^{(\ms)})
~~\forall \ms\in\{1,\dots,\Ms\}.
\end{align}
\end{subequations}
For given constant vectors $\mb\ratevar^{(\ms)}$ and $\mb\rate^{(\ms)}$, this is a linear program in $\mb\mu$ and $z$.
By solving for the optimal $\mb\mu^\star$,
setting $(\mb\ratevar^{(\Ms+1)},
\mb\pow^{(\Ms+1)})
=
(\mb\ratevar^\star(\mb\mu^\star),
\mb\pow^\star(\mb\mu^\star))$,
and incrementing $\Ms$, a refined approximation is obtained.
In every iteration, a feasible approximate solution to the primal problem \eqref{eq:appl:smapel:prob_rewrite} can be recovered by solving the dual linear program of \eqref{eq:appl:smapel:dual_CP}.
These solutions converge from below to the global \chg{optimum \cite[Sec.~6.5]{BaShSh06}}.
Note that primal recovery implicitly performs the convex hull operation corresponding to the rate averaging in \eqref{eq:appl:tsrates} if needed \cite[Sec.~3.3.2]{Br12}.
In addition, each iteration delivers a feasible value of the dual problem in \eqref{eq:appl:smapel:prob_dualsplit},
which acts as an upper bound to the global optimum of \eqref{eq:appl:smapel:prob_rewrite}.
As a termination criterion, we thus check whether the difference of these values is below a predefined accuracy threshold $\cpeps$.

In each iteration of the cutting plane method, evaluating $\mb\ratevar^\star$ and $\mb\pow^\star$ requires solving the inner problems \eqref{eq:appl:smapel:prob_inner}.
The first maximization \eqref{eq:appl:smapel:prob_innerA} is a convex program due to the assumption of a concave utility.
In the special case of the proportional fair utility
\eqref{eq:PFutility}
it can even be solved in closed form.

The challenging nonconvex problem \eqref{eq:appl:smapel:prob_innerB} is a weighted sum rate maximization, which can be tackled by any of the methods discussed in \cref{sec:appl:wsrmax}.
In \cite{Br12,UtBr12}, it was proposed to apply the \cgls{pa} with the rates as optimization variables.
Motivated by the run time comparison from \cref{sec:appl:wsrmax}, we instead use \chg{\cref{alg:bb} together with the reduction procedure from \cref{sec:appl:red}.}
Combining this approach with the cutting plane method for the outer problem, we get the guarantee that the obtained solution lies at most
$\eta+\cpeps$ away from the global optimum, i.e., it is $(\eta+\cpeps)$-optimal.

For a run time comparison using the implementation in \cite{github}, we reconsider the example from \cite{Qian2010} with $K=4$ interfering links and channel gains derived from a path loss model for the network topology given in \cite[Fig.~5]{Qian2010}.
As in \cite{Qian2010}, we maximize the proportional fair utility \eqref{eq:PFutility} without any constraints on the per-user rates.
Since the original S-MAPEL algorithm did not converge within a reasonable amount of time, we use the A-S-MAPEL heuristic with accuracy $\eta=10^{-2}$ and $\asmeps=10^{-3}$.
Unlike S-MAPEL, this accelerated heuristic cannot give a rigorous guarantee for the quality of the obtained solutions  \cite{Qian2010}, but we can use its run time of $3146$~seconds as a (very loose) lower bound for the actual run time of S-MAPEL.
\chg{Instead, the} proposed combination of the cutting plane algorithm and the \cgls{mmp} framework with \chg{total tolerance} of
$\eta+\cpeps=9\cdot 10^{-3}+1\cdot 10^{-3}=10^{-2}$ converged in only $1.77$~seconds.

%

\subsection{Coded Time-Sharing and Rate Balancing}
\label{sec:appl:ratebal}
The combination of a Lagrangian dual approach and the \cgls{mmp} framework can be extended to solve \chg{several other} problems.
For instance, we can consider coded \chg{time-sharing \cite{ElGamal2011}}
where not only the rates but also the transmit powers \chg{are} averaged.
In this case, we have to dualize the resulting average power constraints $\sum_{\ms=1}^\Ms \tau_\ms \pow_k^{(\ms)}\leq P_k$ in addition to the rate constraints.
Moreover, we could replace the fairness optimization by a so-called rate balancing problem, which can be used to characterize the Pareto boundary of the rate region
\cite{MoZhCi06} and to guarantee the quality of service of all users.

As an example, let us combine both mentioned modifications
in an \cgls{ic} under the assumptions of Gaussian inputs, coded time-sharing, and treating interference as noise.
The resulting rate balancing problem with coded time-sharing \chg{is}
\begin{subequations}
\label{eq:appl:ratebal}
\begin{align}
\max_{\substack{(\pow_k^{(\ms)}\geq 0)_{\allk\forall\ms}\\\Ms\in\mathbb{N},R\in\mathbb{R}\\\tau\geq\zero: \one^\Tr\mb\tau=1}} R  \quad\st\quad 
& \sum_{\ms=1}^\Ms \tau_\ms r_k^{(\ms)} \geq \rho_k R,~~\allk \\[-4mm]
& \sum_{\ms=1}^\Ms \tau_\ms \pow_k^{(\ms)}\leq P_k, ~~\allk 
\end{align}
\end{subequations}
for given relative rate targets $\rho_k$, $k=1,\dots,K$,
and with $\rate_k^{(\ms)}$
from \eqref{eq:appl:tsrates}.
After introducing dual variables $\mb\mu$ and $\mb\lambda$ for the rate constraints and power constraints, respectively, 
and performing some reformulations similar to the ones in \cref{sec:appl:scheduling},
the dual problem of \eqref{eq:appl:ratebal} can be written as \cite{HeUt19}
\begin{multline}
\label{eq:appl:ratebal:prob_dualsplitTS}
\min_{\substack{\mb\mu\geq\zero,\mb\lambda\geq\zero\\\mb\rho^\Tr\mb\mu=1}}\Bigg(
\sum_{k=1}^K \lambda_k P_k + 
\max_{(\pow_k\geq 0)_{\allk}} \!\!
(\mb\mu^\Tr \mb\rate  - \mb\lambda^\Tr \mb\pow) 
\Bigg).
\end{multline}
The inner maximization is no longer a pure weighted sum rate problem,
but using the \cgls{mm} function
\begin{equation}
\label{eq:appl:smapel:mmrate}
\brbF(\brbx,\brby) = \sum_{k=1}^K \left(\mu_k \mmrate_k(\brbx,\brby)  - \lambda_k \brbyk[k] \right) 
\end{equation}
it can still be solved via the \cgls{mmp} framework.

In \cite{HeUt19}, this problem was considered for the special case of a two-user \cgls{siso} \cgls{ic},
and a \cgls{bb} solution was proposed for the inner maximization.
In fact, this solution can be considered as a special case of the \cgls{mmp} framework with \cgls{mm} function \eqref{eq:appl:smapel:mmrate}.
For further details and numerical simulations \chg{of} the rate balancing problem, the reader is referred to \cite{HeUt19}.

\subsection{Multiantenna Interference Channels}
The \cgls{mmp} framework can also be used in multiantenna scenarios.
In \chg{a} \cgls{simo} \cgls{ic}
\begin{equation}
\bm y_k = \sum_{\notk=1}^K \bm h_{k \notk} x_\notk + \noisevar_k
\end{equation}
the achievable rates \chg{with Gaussian codebooks, interference treated as noise, and without self-interference} can be expressed \chg{as \cite{YeBl03}}
\begin{equation}
\label{eq:appl:simorate}
\rate_k = \log_2 \Big( 1+ \pow_k \bm h_{kk}^\He \Big(\id_{M_k} + \sum_{\notk \neq k} \pow_\notk \bm h_{k \notk}\bm h_{k \notk}^\He \Big)^{-1} \bm h_{kk} \Big)
\end{equation}
where $M_k$ is the number of antennas at receiver $k$ and $\id_{M_k}$ is the identity matrix of this size.
By replacing the rate \eqref{eq:tinrate} by \eqref{eq:appl:simorate},
we can formulate the weighted sum rate maximization (\cref{sec:appl:wsrmax}), the energy efficiency optimizations (\cref{sec:appl:gee}),
the scheduling problem (\cref{sec:appl:scheduling}), and the rate balancing problem (\cref{sec:appl:ratebal}) for the \cgls{simo} \cgls{ic}.

By calculating the partial derivatives with respect to $\brbx$ and $\brby$ in order to study monotonicity, it can be verified that
\begin{equation}
\label{eq:appl:simorate_mm}
\mmrate_k(\brbx,\brby) = \log_2 \Big( 1+ \brbxk[k] \bm h_{kk}^\He \Big(\id_{M_k} + \sum_{\notk \neq k} \brbyk[\notk] \bm h_{k \notk}\bm h_{k \notk}^\He \Big)^{-1} \bm h_{kk} \Big)
\end{equation}
is a \chg{\cgls{mmp} representation of \cref{eq:appl:simorate}.}
We can thus directly apply the \cgls{mmp} framework to solve all \chg{of} the above-mentioned problems in the \cgls{simo} \cgls{ic}.


For the \cgls{miso} \cgls{ic} with multiple antennas at the transmitter side,
the optimization is more involved since transmit covariance matrices or beamforming vectors need to be designed instead of transmit powers.
In the following, we present a beamformer-based method for the two-user \cgls{miso} \cgls{ic}
\begin{equation}
y_k = \mb h_{kk}^\He \mb x_k + \mb h_{k\notk}^\He \mb x_\notk + \noise_k
\end{equation}
with $\notk=3-k$.
The transmit signals $\mb x_k = \sqrt{\pow_k} \mb b_k s_k$ are generated from scalar Gaussian inputs $s_k\sim\mathcal{CN}(0,1)$,
where $\mb b_k$ is a normalized beamforming vector with $\|\mb b_k\|=1$.

The approach is based on \cite{Jorswieck2010a}, which uses parameters $\misoparam_k$ to construct a convex combination of the \gls{mrt} beamformer and the \gls{zf} beamformer,
which is provably sufficient to parameterize all Pareto-optimal transmit strategies in the considered scenario.
We thus use the beamforming vectors
\begin{subequations}
\begin{align}
\mb b_k &= \nofracX{\tilde{\mb b}_k}{\|\tilde{\mb b}_k\|}\!,\!\! & \tilde{\mb b}_k&= \zeta_k \mb b_k^\text{MRT} + (1-\zeta_k) \mb b_k^\text{ZF},\!\!
\\
\!\mb b_k^\text{MRT} &= \nofracX{\mb h_{kk}}{\|\mb h_{kk}\|}\!,\!\! &
\!\mb b_k^\text{ZF} &= \nofracX{\mb\Pi_{\mb h_{\notk k}}^\bot \mb h_{kk}}{\|\mb\Pi_{\mb h_{\notk k}}^\bot \mb h_{kk}\|} \!
\end{align}
\end{subequations}
where $\mb\Pi_{\mb h_{\notk k}}^\bot = \id_{M_k}-\frac{\mb h_{\notk k}\mb h_{\notk k}^\He}{\mb h_{\notk k}^\He\mb h_{\notk k}}$
is the orthogonal projection onto the orthogonal complement of the span of $\mb h_{\notk k}$,
and $\zeta_k$, $k=1,2$ are auxiliary variables that need to be optimized.
The achievable rates \chg{with Gaussian codebooks, interference treated as noise, and without self-interference} can then be expressed \chg{as \cite{Jorswieck2010a}}
\begin{align}
r_k \!=\! \log_2\bigg(\!1 + \frac{ \pow_k |\mb h_{kk}^\He \mb b_k|^2 }{\sigma_k^2 + \pow_\notk|\mb h_{k\notk}^\He \mb b_\notk|^2} \!\bigg)
\!=\!\log_2\left(\!1+ \frac{\pow_k \alpha_k(\mb\zeta)}{ \sigma^2 + \pow_\notk \beta_\notk(\mb\zeta) }\!\right) 
\label{eq:appl:misorate}
\end{align}
where
\begin{subequations}
\begin{align}
\alpha_k(\mb\zeta) &= |\mb h_{kk}^\He \mb b_k|^2 
= \frac{(\zeta_k \gamma_{kk} + (1-\zeta_k)\gamma_{k\notk})^2 }{1- 2\zeta_k(1-\zeta_k)(1-\frac{\gamma_{k\notk}}{\gamma_{kk}})} \geq 0, 
\\
\beta_k(\mb\zeta) &=  |\mb h_{k\notk}^\He \mb b_\notk|^2
= \frac{\zeta_k^2 \delta_{k\notk}^2  \gamma_{kk}^{-2}}{1- 2\zeta_k(1-\zeta_k)(1-\frac{\gamma_{k\notk}}{\gamma_{kk}})} \geq 0
\end{align}
\end{subequations}
with $\gamma_{kk} = \|\mb h_{kk}\|$, 
$\gamma_{k\notk} = \|\mb\Pi_{\mb h_{\notk k}}^\bot \mb h_{kk}\|$,
and $\delta_{k\notk} = |\mb h_{kk}^\He \mb h_{k\notk} |$
\cite{Jorswieck2010a,HeMaJoUt19}.
Since $\alpha_k(\mb\zeta)$ and $\beta_k(\mb\zeta)$ are nondecreasing in both components of $\mb\zeta$ (see \cite{Jorswieck2010a} for a proof),
we can use \cref{eq:mmrules:sum,eq:mmrules:prod,eq:mmrules:incdec} to establish the \cgls{mmp} rate expression
\begin{equation}
\label{eq:appl:miso_mmrate}
\mmrate_k\left(
{\scriptsize\begin{bmatrix}\bm\zeta \\ \bm p\end{bmatrix}},
{\scriptsize\begin{bmatrix}\bm\xi \\ \bm q\end{bmatrix}}
\right) = \log_2\left(1+ \frac{\pow_k \alpha_k(\bm\zeta)}{ \sigma^2 + q_\notk \beta_\notk(\bm\xi) }\right).
\end{equation}

We can thus optimize the global energy efficiency in the two-user \cgls{miso} \cgls{ic}
by replacing $\mmrate_k$ in \eqref{eq:appl:gee_mm} by \eqref{eq:appl:miso_mmrate}
with $\brbx=[\bm\zeta^\Tr, \bm p^\Tr]^\Tr$ and $\brby=[\bm\xi^\Tr, \bm q^\Tr]^\Tr$.
This means that we apply \chg{\cref{alg:bb}} in a four-dimensional space.

In a similar manner, all other optimization problems for the single-antenna interference channel that could previously be formulated by means of the \cgls{mmp} rate expression $\mmrate_k$ from \eqref{eq:tinrate_mm}
can be easily extended to the two-user \cgls{miso} interference channel by using the \cgls{mmp} rate expression \eqref{eq:appl:miso_mmrate} instead.
An example for this is the rate balancing problem \eqref{eq:appl:ratebal} in the \cgls{miso} \cgls{ic} which we considered in \cite{HeMaJoUt19}.
For the special case of weighted sum rate maximization without minimum rate constraints, the problem can be simplified since,
\chg{in this case, it is optimal for both users to exploit} their full power budget \cite[Proposition~1]{LaJo08}.
\chg{Hence, \cref{alg:bb} has to be applied only for the two auxiliary variables $\bm\zeta$.}

The \cgls{mmp} framework can also be applied to nonconvex optimization problems in other multiantenna scenarios, \chg{such as the} $K$-user \cgls{miso} broadcast channel with linear transceivers.
An example is the method in \cite[Sec.~7.3.1.2]{He17}, which is in fact a special case of the \cgls{mmp} framework.

\subsection{Probability Optimization for Slotted ALOHA} \label{sec:aloha}
To demonstrate that the proposed \cgls{mmp} framework can also be useful for solving problems on the medium access control layer,
we study the problem from \cite[Ch.~7]{Zhang2012} where the transmission probabilities in the slotted ALOHA protocol with $K$ users were optimized, i.e., 
\begin{subequations}
\label{eq:appl:aloha:orig}
\begin{align}
\max_{\zero\leq\mb\alohaP\leq\one}~
& \utility \left( \rate_1(\mb\alohaP),\dots,\rate_K(\mb\alohaP)   \right)
\\\st~~ &\rate_k(\mb\alohaP) \geq \rmink ~~\allk
\end{align}
\end{subequations}
with an increasing (not necessarily concave) utility function $\utility$, and average per-user throughput
\begin{equation}
\rate_k(\mb\alohaP) = \alohaR_k \alohaP_k \prod\nolimits_{\notk \in \mathcal{I}(k)} (1- \alohaP_\notk).
\end{equation}
Here, $\mb\alohaP=[\alohaP_1,\dots,\alohaP_K]^\Tr$ contains the probabilities $\alohaP_k$ that user $k$ attempts to transmit a packet in any time-slot,
and $\mathcal{I}(k)$ contains the indices of all users that cause interference to receiver $k$.
The data rates $\rate_k$ are given by the product of the data rate $\alohaR_k$ of a successful transmission and the probability of a collision-free transmission.

The first solution approach in \cite[Ch.~7]{Zhang2012} transforms the problem to a canonical monotonic optimization problem
\begin{subequations}
\label{eq:appl:aloha:double}
\begin{align}
\max_{\mb\alohaP\geq\zero,\hat{\mb\alohaP}\geq\zero} ~
&\utility \left( \hat\rate_1(\mb\alohaP,\hat{\mb\alohaP}),\dots,\hat\rate_K(\mb\alohaP,\hat{\mb\alohaP})   \right)
\\\st~~ &\hat\rate_k(\mb\alohaP,\hat{\mb\alohaP}) \geq \rmink ~~\allk
\\ &\mb\alohaP+\hat{\mb\alohaP} \leq \one
\end{align}
\end{subequations}
with
\begin{equation}
\hat\rate_k(\mb\alohaP,\hat{\mb\alohaP}) = \alohaR_k \alohaP_k \prod\nolimits_{\notk \in \mathcal{I}(k)} \hat\alohaP_\notk
\end{equation}
and solves it by means of the \cgls{pa}.
As an alternative, this problem could also be solved with the \cgls{brb} algorithm for \cgls{dm} problems from \cite[Sec.~7]{Tuy2005}.
However, no matter which algorithm is applied, the formulation in \eqref{eq:appl:aloha:double} suffers from the doubled dimensionality of the optimization problem,
which has drastic consequences \cite[Ch.~7]{Zhang2012} since the worst-case complexities of the \cgls{pa} and \chg{\cgls{bb}} algorithm grows exponentially in the number of variables \cite{Bjornson2013}.

Therefore, a second approach
\begin{equation}
\label{eq:appl:aloha:gp}
\max_{\mb\alohaY\geq\zero} ~
\utility \left( \alohaR_1 \alohaY_1,,\dots,\alohaR_K \alohaY_K  \right) ~~\st~~ \mb\alohaY\in\alohaYset
\end{equation}
was proposed in \cite[Ch.~7]{Zhang2012}, where
\begin{multline}
\alohaYset = \big\{ \mb\alohaY \,|\, \alohaR_k\alohaY_k\geq \rmink,~\allk ~\text{and}~ \\
\exists (\zero\leq\mb\alohaP\leq\one): \alohaR_k\alohaY_k=\rate_k(\mb\alohaP),~\allk 
\big\}.
\end{multline}
As a result, the \cgls{pa} algorithm can be implemented with \chg{only} $K$ variables,
but this comes at the cost that
a geometric program (for details see \cite[Ch.~7]{Zhang2012}) has to be solved 
to perform the projection to $\mb\alohaY\in\alohaYset$ in each iteration of the \cgls{pa}.

To avoid the drawbacks of both methods, we reformulate \eqref{eq:appl:aloha:orig}
in terms of the \cgls{mmp} framework with \cgls{mm} objective and \cgls{mm} constraint functions given as
\begin{subequations}
\label{eq:appl:aloha:mmp}
\begin{align}
\brbF(\brbx,\brby) &=  \utility \left( \Rate_1(\brbx,\brby),\dots,\Rate_K(\brbx,\brby)   \right), \\
\brbG_k(\brbx,\brby) &= \rmink - \Rate_k(\brby,\brbx), \\
\text{with~} \Rate_k(\brbx,\brby) &= \alohaR_k \brbxk[k] \prod\nolimits_{\notk \in \mathcal{I}(k)} (1- \brbyk[\notk]).
\end{align}
\end{subequations}

Note that these constraints do not fulfill the additional requirements in \cref{mmp:feas:prop}. Thus, \cref{feas:mmp} is not applicable and \cref{eq:appl:aloha:orig} with \cgls{mmp} representations \cref{eq:appl:aloha:mmp} needs to be solved with the modified, infinite version of \cref{alg:bb} described in \cref{sec:feasibility}. However, although the algorithm is infinite in theory, it turns out to have very fast convergence in practice.


It is important to note that the auxiliary variables $\brby$ in the \cgls{mmp} method are used only as a vehicle to compute bounds,
without considering them as additional optimization variables.
Thus, unlike the canonical monotonic reformulation \eqref{eq:appl:aloha:double},
the \cgls{mmp} method does not increase the dimensionality of the problem.
Moreover, the \cgls{mmp} formulation avoids an additional inner solver as needed in the \chg{geometric programming based} formulation \eqref{eq:appl:aloha:gp}.

For the numerical results in \cref{tab:appl:aloha:runtime}, we have used a three-user system with proportional fair utility \eqref{eq:PFutility}, full interference $\mathcal{I}(k)=\{\notk\,|\,\notk\neq k\}$,
and $\alohaR_k=\log_2(1+\abs{\alpha_k'}^2)$ with \cgls{iid} $\alpha_k'\sim \mathcal{CN}(0, 1)$ for all $k$.
To create a variety of challenging scenarios in which some of the minimum rate constraints are active, we have generated $\rmink=\alohaR_k \chi_k$ with $\chi_k\sim\mathcal{N}\big(\frac{(K-1)^2}{K^2}, .05^2\big)$
since $\rmink=\frac{(K-1)^2}{K^2}\alohaR_k,~\allk$ would be the boundary to infeasibility in case of full interference.
All infeasible scenarios among the generated ones have been discarded, and the results are averaged over $100$ feasible scenarios.
As all algorithms have fundamentally different per-iteration complexities, it does not make sense to count iterations.
We thus again fall back to comparing computation times of the C++ implementations \cite{github}.

\begin{table}
	\caption{Mean and median run times of various solution methods for \eqref{eq:appl:aloha:orig}.}
\label{tab:appl:aloha:runtime}
\addtolength\tabcolsep{-1pt}
\centering
\begin{tabular}{lrrrr}
	\toprule
& \multicolumn{2}{c}{$3$ Users} & \multicolumn{2}{c}{$4$ Users}
\\
& Mean & Median & Mean & Median
\\
\midrule
\eqref{eq:appl:aloha:double} \& \cgls{pa} & \multicolumn{2}{c}{$> \SI{23}{\hour}$} & \multicolumn{1}{c}{---} & \multicolumn{1}{c}{---}
\\
\eqref{eq:appl:aloha:double} \& \chg{\cgls{dm} \cgls{bb}} (no reduction) & \SI{88.250}{\second} & \SI{21.305}{\second} & \multicolumn{1}{c}{---}  & \multicolumn{1}{c}{---}
\\
\eqref{eq:appl:aloha:double} \& \chg{\cgls{dm} \cgls{brb}} (reduction) & \SI{23.182}{\second} & \SI{5.919}{\second} & \multicolumn{1}{c}{---}  & \multicolumn{1}{c}{---}
\\
\eqref{eq:appl:aloha:gp} \& \cgls{pa} & \SI{3.629}{\second} & \SI{0.961}{\second} & \SI{13.4}{\hour} & \SI{22.935}{\second} 
\\
\cgls{mmp} \eqref{eq:appl:aloha:mmp} (no reduction) & \SI{0.769}{\second} & \SI{0.172}{\second} & \SI{150.1}{\second} & \SI{4.838}{\second} 
\\
\cgls{mmp} \eqref{eq:appl:aloha:mmp} (reduction)& \SI{1.413}{\second} & \SI{0.364}{\second} & \SI{256.0}{\second} & \SI{8.139}{\second} 
\\
\bottomrule
\end{tabular}\vspace*{-3mm}
\end{table}

In addition to the significantly lower run time of the \cgls{mmp} method, another remarkable aspect can be observed.
The reduction step \eqref{eq:reduction:rs} reduces the run time of the \cgls{dm} approach while it increases the run time of the \cgls{mmp} approach.
As stated before, it depends on the problem under consideration whether or not performing a reduction leads to an overall gain in computation time.
An example where the reduction step proves to be very helpful in combination with the \cgls{mmp} method is the rate balancing problem with time-sharing in \cite{HeMaJoUt19}.

\section{Discussion}
\label{sec:discuss}
{\glsreset{mmp}
\glsreset{mm}

The \cgls{mmp} framework that we propose in this article can directly exploit hidden monotonicity of single terms in a function expression
even if the function as a whole is neither monotonic nor a difference of monotonic functions.
This allows us to derive bounds that are tighter than previously used \cgls{dm} bounds,
leading to faster \chg{convergence in \cgls{bb} algorithms}.
Moreover, the \cgls{mmp} framework enables us to derive bounds even for a wide range of problems for which no \cgls{dm} reformulation exists,
so that we can avoid previously proposed nested algorithms, e.g., for fractional monotonic problems.
Due to these advantages, solutions based on the new \cgls{mmp} framework achieve tremendous reductions of run time and memory consumption compared to state-of-the-art solutions
in all numerical examples that we considered.
\chg{These examples come from the area of signal processing for communications, but we are convinced that the proposed framework can help to speed up global optimization in many other areas of research as well.}

An interesting theoretical aspect of the \cgls{mmp} framework is that it can be considered as a generalization of the \cgls{dm} approach and of other special cases previously studied in the literature.
From a practical perspective, we have discussed the oldest-first selection rule and a reduction method for \cgls{mmp} problems
as additional methods to speed up the implementation in specific scenarios.
In the code repository \cite{github}, we provide a C++ implementation of the proposed \cgls{brb} algorithm for \cgls{mmp},
which can be easily adapted to arbitrary \cgls{mmp} problems,
as well as the simulation code for all numerical examples discussed in this paper.

\chg{\subsection{Convergence Speed and the Optimal Choice of $F$}
\label{sec:discuss:convergence}
	We have established convergence of the \cGls{brb} algorithm for \cgls{mmp} problems (\cref{alg:bb}) to an $\eta$-optimal solution of \cref{eq:MMP} within a finite number of iterations  in \cref{thm:bbconv}. Establishing this kind of convergence is an important theoretical result as pointed out by Donald Knuth \cite[Sect.~1.1]{Knuth1997vol1}. In case of \cref{alg:bb}, it holds for any \cgls{mmp} representation $F$ of $f$, i.e., any continuous function $F$ that satisfies \cref{mmp:prop12,mmp:prop3}. Thus, from a theoretical perspective, the non-uniqueness of \cgls{mmp} representations is not an issue for the convergence of \cref{alg:bb}.
	
	However, we have seen in \cref{sec:appl} that the actual convergence speed depends strongly on the precise choice of $F$.
	It would be beneficial to have means of obtaining the optimal $F$ for a given objective function $f$. This raises the question of how to define \emph{optimality} in terms of a bounding function. From a theoretical perspective, the tightest bound leads to fastest convergence. Practically, however, this bound can be costly to compute and might lead to longer run times than less tight bounds. In some problem instances, a tighter bound might even lead to slower convergence because the branching is performed in a different order and better feasible points are encountered earlier. 
	
	Even when leaving all these considerations aside and simply assuming that the optimal $F$ is the \cgls{mmp} representation of $f$ that leads to the tightest bound, it is still challenging to obtain general quantitative statements that are not limited to a particular problem instance. Indeed, obtaining tight (not even the tightest) bounds is not only relevant for \cgls{mmp}, or monotonic optimization in general, but also in other fields of nonconvex optimization, both for global optimal solutions and heuristic algorithms \cite{Sun2017a}. For example, every successive convex approximation algorithm benefits from tight bounds, but in many cases it is already a noticeable achievement to obtain \emph{some} suitable bound. This is where one of our main contributions lies: We provide the theoretical and algorithmic framework for a novel bounding methodology that enables the derivation of powerful bounds for functions that were previously intractable (cf.~\cref{sec:appl:gee}).

	Another aspect is comparing the convergence behavior of different algorithms (e.g., \cgls{pa} and \cgls{mmp}) and bounding schemes (e.g., \cgls{dm} and \cgls{mmp}) analytically. The common methodology is to examine the rate of convergence defined as the number $p\in\mathds N$ such that
	\begin{equation}
	\label{eq:convergence_rate}
	U(\mathcal M_k) - \gamma_k \le C \diam(\mathcal M_k)^p
	\end{equation}
	for some fixed constant $C > 0$. This is an active area of research in the operations research community and leads to worst-case bounds on the number of required iterations. However, supported by the results in \cite{Schoebel2009}, we would expect a rate of convergence $p \le 2$ for \cgls{dm} and \cgls{mmp} bounds. Moreover, we expect this rate to be equal for both bounding schemes since \cgls{dm} bounds are a special case of \cgls{mmp} bounds. Thus, supporting the experimental results in \cref{sec:appl} by theoretical results would require an approach that goes beyond the usual rate of convergence analysis and determines the constant $C$ in \eqref{eq:convergence_rate} analytically. Again, the challenge would be to obtain quantitative statements that are not specific to a particular problem instance. Moreover, since such results will only be worst-case bounds without direct implications for the average run time, it is not clear whether this approach is a viable method to quantify the experimentally verified performance gain of \cgls{mmp}. For all these reasons, such an analysis goes beyond the scope of this paper and is left open for future research.
}

\chg{Apart from these theoretical questions, we have demonstrated
that the \cgls{mmp} framework helps to find much faster globally optimal solution methods for many relevant problems.
This can be observed numerically from the broad selection of examples in \cref{sec:appl}, it can be justified analytically in some cases such as in \cref{sec:appl:wsrmax}, and it is obvious from the fact that the \cgls{mmp} framework helps to avoid nested optimization in other cases.
Especially in cases where a \cgls{dm} representation is not available but an \cgls{mmp} representation can be found, \cgls{mmp} leads to tremendous speed-ups over the state-of-the-art, even though we might not have a theoretical guarantee to have found the optimal \cgls{mmp} representation. Moreover, in some of these cases it even enables the solution of previously intractable problems.}
}



\balance
\bibliography{IEEEabrv,Abrv,paper}

\begin{thebibliography}{10}
\providecommand{\url}[1]{#1}
\csname url@samestyle\endcsname
\providecommand{\newblock}{\relax}
\providecommand{\bibinfo}[2]{#2}
\providecommand{\BIBentrySTDinterwordspacing}{\spaceskip=0pt\relax}
\providecommand{\BIBentryALTinterwordstretchfactor}{4}
\providecommand{\BIBentryALTinterwordspacing}{\spaceskip=\fontdimen2\font plus
\BIBentryALTinterwordstretchfactor\fontdimen3\font minus
  \fontdimen4\font\relax}
\providecommand{\BIBforeignlanguage}[2]{{%
\expandafter\ifx\csname l@#1\endcsname\relax
\typeout{** WARNING: IEEEtran.bst: No hyphenation pattern has been}%
\typeout{** loaded for the language `#1'. Using the pattern for}%
\typeout{** the default language instead.}%
\else
\language=\csname l@#1\endcsname
\fi
#2}}
\providecommand{\BIBdecl}{\relax}
\BIBdecl

\bibitem{MaHeJo19}
B.~Matthiesen, C.~Hellings, and E.~A. Jorswieck, ``Energy efficiency: Rate
  splitting vs. point-to-point codes in {Gauss}ian interference channels,'' in
  \emph{Proc. IEEE Int. Workshop Signal Process. Adv. Wireless Commun.
  (SPAWC)}, Cannes, France, Jul. 2019.

\bibitem{HeMaJoUt19}
C.~Hellings, B.~Matthiesen, E.~A. Jorswieck, and W.~Utschick, ``Globally
  optimal {TIN} strategies with time-sharing in the {MISO} interference
  channel,'' in \emph{Proc. Eur. Signal Process. Conf. (EUSIPCO)}, A
  Coru{\~n}a, Spain, Sep. 2019.

\bibitem{Te99}
I.~E. Telatar, ``Capacity of multi-antenna {G}aussian channels,'' \emph{Eur.
  Trans. Telecommun.}, vol.~10, no.~6, pp. 585--595, Nov./Dec. 1999.

\bibitem{PaCiLa03}
D.~Palomar, J.~Cioffi, and M.~Lagunas, ``Joint {Tx-Rx} beamforming design for
  multicarrier {MIMO} channels: a unified framework for convex optimization,''
  \emph{{IEEE} Trans. Signal Process.}, vol.~51, no.~9, pp. 2381--2401, Sep.
  2003.

\bibitem{Isheden2012}
C.~Isheden, Z.~Chong, E.~Jorswieck, and G.~Fettweis, ``Framework for link-level
  energy efficiency optimization with informed transmitter,'' \emph{{IEEE}
  Trans. Wireless Commun.}, vol.~11, no.~8, pp. 2946--2957, Aug. 2012.

\bibitem{Jindal2005}
N.~Jindal, W.~Rhee, S.~Vishwanath, S.~A. Jafar, and A.~Goldsmith, ``Sum power
  iterative water-filling for multi-antenna {G}aussian broadcast channels,''
  \emph{{IEEE} Trans. Inf. Theory}, vol.~51, no.~4, pp. 1570--1580, Apr. 2005.

\bibitem{Liu2012}
L.~Liu, R.~Zhang, and K.-C. Chua, ``Achieving global optimality for weighted
  sum-rate maximization in the {K}-user {Gauss}ian interference channel with
  multiple antennas,'' \emph{{IEEE} Trans. Wireless Commun.}, vol.~11, no.~5,
  pp. 1933--1945, May 2012.

\bibitem{YeBl03}
S.~Ye and R.~S. Blum, ``Optimized signaling for {MIMO} interference systems
  with feedback,'' \emph{{IEEE} Trans. Signal Process.}, vol.~51, no.~11, pp.
  2839--2848, Nov. 2003.

\bibitem{BoKa08}
R.~B\"ohnke and K.-D. Kammeyer, ``Weighted sum rate maximization for
  {MIMO-OFDM} systems with linear and dirty paper precoding,'' in \emph{Proc.
  Int. ITG Conf. Source, Channel Coding (SCC)}, Ulm, Germany, Jan. 2008.

\bibitem{HuScJo08}
R.~Hunger, D.~Schmidt, and M.~Joham, ``A combinatorial approach to maximizing
  the sum rate in the {MIMO BC} with linear precoding,'' in \emph{Proc.
  Asilomar Conf. Signals, Syst., Comput.}, Pacific Grove, CA, USA, Oct. 2008,
  pp. 316--320.

\bibitem{HeUtJo11}
C.~Hellings, W.~Utschick, and M.~Joham, ``Power minimization in parallel vector
  broadcast channels with separate linear precoding,'' in \emph{Proc. Eur.
  Signal Process. Conf. (EUSIPCO)}, Barcelona, Spain, Sep. 2011, pp.
  1834--1838.

\bibitem{YoGo06}
T.~Yoo and A.~Goldsmith, ``On the optimality of multiantenna broadcast
  scheduling using zero-forcing beamforming,'' \emph{{IEEE} J. Sel. Areas
  Commun.}, vol.~24, no.~3, pp. 528--541, Mar. 2006.

\bibitem{GuUtHuJo10}
C.~Guthy, W.~Utschick, R.~Hunger, and M.~Joham, ``Efficient weighted sum rate
  maximization with linear precoding,'' \emph{{IEEE} Trans. Signal Process.},
  vol.~58, no.~4, pp. 2284--2297, Apr. 2010.

\bibitem{Zappone2016}
A.~Zappone, L.~Sanguinetti, G.~Bacci, E.~A. Jorswieck, and M.~Debbah,
  ``Energy-efficient power control: A look at {5G} wireless technologies,''
  \emph{{IEEE} Trans. Signal Process.}, vol.~64, no.~7, pp. 1668--1683, 4 2016.

\bibitem{Yang2017}
Y.~Yang and M.~Pesavento, ``A unified successive pseudoconvex approximation
  framework,'' \emph{{IEEE} Trans. Signal Process.}, vol.~65, no.~13, pp.
  3313--3328, Jul. 2017.

\bibitem{CoToJuLa07}
M.~Codreanu, A.~T\"{o}lli, M.~Juntti, and M.~Latva-aho, ``Joint design of
  {Tx-Rx} beamformers in {MIMO} downlink channel,'' \emph{{IEEE} Trans. Signal
  Process.}, vol.~55, no.~9, pp. 4639--4655, Sep. 2007.

\bibitem{ShScBo08}
S.~Shi, M.~Schubert, and H.~Boche, ``Rate optimization for multiuser {MIMO}
  systems with linear processing,'' \emph{{IEEE} Trans. Signal Process.},
  vol.~56, no.~8, pp. 4020--4030, Aug. 2008.

\bibitem{ChAgCaCi08}
S.~S. Christensen, R.~Agarwal, E.~D. Carvalho, and J.~M. Cioffi, ``Weighted
  sum-rate maximization using weighted {MMSE} for {MIMO-BC} beamforming
  design,'' \emph{{IEEE} Trans. Wireless Commun.}, vol.~7, no.~12, pp.
  4792--4799, Dec. 2008.

\bibitem{LaAgVi16}
S.~Lagen, A.~Agustin, and J.~Vidal, ``Coexisting linear and widely linear
  transceivers in the {MIMO} interference channel,'' \emph{{IEEE} Trans. Signal
  Process.}, vol.~64, no.~3, pp. 652--664, Feb. 2016.

\bibitem{ScShBeHoUt09}
D.~A. Schmidt, C.~Shi, R.~A. Berry, M.~L. Honig, and W.~Utschick, ``Pricing
  algorithms for power control and beamformer design in interference
  networks,'' \emph{{IEEE} Signal Process. Mag.}, vol.~26, no.~5, pp. 53--63,
  Sep. 2009.

\bibitem{LaJo08}
E.~G. Larsson and E.~A. Jorswieck, ``Competition versus cooperation on the
  {MISO} interference channel,'' \emph{{IEEE} J. Sel. Areas Commun.}, vol.~26,
  no.~7, pp. 1059--1069, Sep. 2008.

\bibitem{ScPaBa08a}
G.~Scutari, D.~P. Palomar, and S.~Barbarossa, ``Optimal linear precoding
  strategies for wideband noncooperative systems based on game theory---part
  {II}: Algorithms,'' \emph{{IEEE} Trans. Signal Process.}, vol.~56, no.~3, pp.
  1250--1267, Mar. 2008.

\bibitem{Matthiesen2015}
B.~Matthiesen, A.~Zappone, and E.~A. Jorswieck, ``Resource allocation for
  energy-efficient 3-way relay channels,'' \emph{{IEEE} Trans. Wireless
  Commun.}, vol.~14, no.~8, pp. 4454--4468, Aug. 2015.

\bibitem{Baccelli2011}
F.~Baccelli, N.~Bambos, and N.~Gast, ``Distributed delay-power control
  algorithms for bandwidth sharing in wireless networks,'' \emph{{IEEE/ACM}
  Trans. Netw.}, vol.~19, no.~5, pp. 1458--1471, Oct. 2011.

\bibitem{Horst1996}
R.~Horst and H.~Tuy, \emph{Global Optimization: Deterministic Approaches},
  3rd~ed.\hskip 1em plus 0.5em minus 0.4em\relax New York; Berlin, Germany;
  Vienna, Austria: Springer-Verlag, 1996.

\bibitem{Tuy2000}
H.~Tuy, ``Monotonic optimization: Problems and solution approaches,''
  \emph{SIAM J. Optim.}, vol.~11, no.~2, pp. 464--494, 2 2000.

\bibitem{Tuy2005}
H.~Tuy, F.~Al-Khayyal, and P.~T. Thach, ``Monotonic optimization: Branch and
  cut methods,'' in \emph{Essays and Surveys in Global Optimization}, C.~Audet,
  P.~Hansen, and G.~Savard, Eds.\hskip 1em plus 0.5em minus 0.4em\relax New
  York; Berlin, Germany; Vienna, Austria: Springer-Verlag, 2005, ch.~2, pp.
  39--78.

\bibitem{Tuy2016}
H.~Tuy, \emph{Convex Analysis and Global Optimization}, 2nd~ed., ser. Springer
  Optim. Appl.\hskip 1em plus 0.5em minus 0.4em\relax New York; Berlin,
  Germany; Vienna, Austria: Springer-Verlag, 2016, vol. 110.

\bibitem{JoLa08}
E.~A. Jorswieck and E.~G. Larsson, ``Linear precoding in multiple antenna
  broadcast channels: Efficient computation of the achievable rate region,'' in
  \emph{Proc. Int. ITG Workshop Smart Antennas (WSA)}, Darmstadt, Germany, Feb.
  2008, pp. 21--28.

\bibitem{Jorswieck2010a}
------, ``Monotonic optimization framework for the two-user {MISO} interference
  channel,'' \emph{{IEEE} Trans. Commun.}, vol.~58, no.~7, pp. 2159--2168, Jul.
  2010.

\bibitem{Qian2009}
L.~P. Qian, Y.~J. Zhang, and J.~Huang, ``{MAPEL}: Achieving global optimality
  for a non-convex wireless power control problem,'' \emph{{IEEE} Trans.
  Wireless Commun.}, vol.~8, no.~3, pp. 1553--1563, Mar. 2009.

\bibitem{Qian2010}
L.~P. Qian and Y.~J. Zhang, ``{S-MAPEL}: Monotonic optimization for non-convex
  joint power control and scheduling problems,'' \emph{{IEEE} Trans. Wireless
  Commun.}, vol.~9, no.~5, pp. 1708--1719, May 2010.

\bibitem{Zhang2012}
Y.~J. Zhang, L.~P. Qian, and J.~Huang, \emph{Monotonic Optimization in
  Communication and Networking Systems}, ser. FnT Netw.\hskip 1em plus 0.5em
  minus 0.4em\relax Boston, MA, USA: Now, 2012, vol.~7, no.~1.

\bibitem{Zappone2015}
A.~Zappone and E.~A. Jorswieck, \emph{Energy Efficiency in Wireless Networks
  via Fractional Programming Theory}, ser. FnT Commun. Inf. Theory.\hskip 1em
  plus 0.5em minus 0.4em\relax Boston, MA, USA: Now, 2015, vol.~11, no. 3--4.

\bibitem{Zappone2017}
A.~Zappone, E.~Bj\"{o}rnson, L.~Sanguinetti, and E.~A. Jorswieck, ``Globally
  optimal energy-efficient power control and receiver design in wireless
  networks,'' \emph{{IEEE} Trans. Signal Process.}, vol.~65, no.~11, pp.
  2844--2859, 6 2017.

\bibitem{UtBr12}
W.~Utschick and J.~Brehmer, ``Monotonic optimization framework for coordinated
  beamforming in multicell networks,'' \emph{{IEEE} Trans. Signal Process.},
  vol.~60, no.~4, pp. 1899--1909, Apr. 2012.

\bibitem{Br12}
J.~Brehmer, \emph{Utility Maximization in Nonconvex Wireless Systems}, ser.
  Found. Signal Process., Commun., Netw., W.~Utschick, H.~Boche, and R.~Mathar,
  Eds.\hskip 1em plus 0.5em minus 0.4em\relax New York; Berlin, Germany;
  Vienna, Austria: Springer-Verlag, 2012, vol.~5.

\bibitem{BrUt09}
J.~Brehmer and W.~Utschick, ``Utility maximization in the multi-user {MISO}
  downlink with linear precoding,'' in \emph{Proc. IEEE Int. Conf. Commun.
  (ICC)}, Dresden, Germany, Jun. 2009.

\bibitem{HeJoRiUt12}
C.~Hellings, M.~Joham, M.~Riemensberger, and W.~Utschick, ``Minimal transmit
  power in parallel vector broadcast channels with linear precoding,''
  \emph{{IEEE} Trans. Signal Process.}, vol.~60, no.~4, pp. 1890--1898, Apr.
  2012.

\bibitem{HeUt12}
C.~Hellings and W.~Utschick, ``Energy-efficient rate balancing in vector
  broadcast channels with linear transceivers,'' in \emph{Proc. Int. Symp.
  Wireless Commun. Syst. (ISWCS)}, Paris, France, Aug. 2012, pp. 1044--1048.

\bibitem{Bjoernson2012}
E.~Bj\"{o}rnson, G.~Zheng, M.~Bengtsson, and B.~Ottersten, ``Robust monotonic
  optimization framework for multicell {MISO} systems,'' \emph{{IEEE} Trans.
  Signal Process.}, vol.~60, no.~5, pp. 2508--2523, Jan. 2012.

\bibitem{GrJoUt12}
A.~Gr\"{u}ndinger, M.~Joham, and W.~Utschick, ``Feasibility test and globally
  optimal beamformer design in the satellite downlink based on instantaneous
  and ergodic rates,'' in \emph{Proc. Int. ITG Workshop Smart Antennas (WSA)},
  Dresden, Germany, Mar. 2012, pp. 217--224.

\bibitem{Bjornson2013}
E.~Bj\"{o}rnson and E.~A. Jorswieck, \emph{Optimal Resource Allocation in
  Coordinated Multi-Cell Systems}, ser. FnT Commun. Inf. Theory.\hskip 1em plus
  0.5em minus 0.4em\relax Boston, MA, USA: Now, 2013, vol.~9, no. 2--3.

\bibitem{github}
\BIBentryALTinterwordspacing
B.~Matthiesen and C.~Hellings. (2019) Accompanying source code. [Online].
  Available: \url{https://github.com/bmatthiesen/mixed-monotonic}
\BIBentrySTDinterwordspacing

\bibitem{HeUt18}
C.~Hellings and W.~Utschick, ``Proper time-sharing as a baseline for studying
  improper signaling in interference channels,'' in \emph{Proc. Int. ITG
  Workshop Smart Antennas (WSA)}, Bochum, Germany, Mar. 2018.

\bibitem{HeUt19}
\BIBentryALTinterwordspacing
------, ``Improper signaling versus time-sharing in the two-user {G}aussian
  interference channel with {TIN},'' 2018, in review{IEEE} Trans. Inf. Theory.
  [Online]. Available: \url{https://arxiv.org/abs/1808.01611}
\BIBentrySTDinterwordspacing

\bibitem{dinkelbach1967}
W.~Dinkelbach, ``On nonlinear fractional programming,'' \emph{Manage. Sci.},
  vol.~13, no.~7, pp. 492--498, 3 1967.

\bibitem{diss}
\BIBentryALTinterwordspacing
B.~Matthiesen, ``Efficient globally optimal resource allocation in wireless
  interference networks,'' Ph.D. Thesis, Technische Universit\"at Dresden,
  Dresden, Germany, Nov. 2019. [Online]. Available:
  \url{https://nbn-resolving.org/urn:nbn:de:bsz:14-qucosa2-362878}
\BIBentrySTDinterwordspacing

\bibitem{Tuy1998}
H.~Tuy, \emph{Convex Analysis and Global Optimization}.\hskip 1em plus 0.5em
  minus 0.4em\relax Kluwer, 1998.

\bibitem{Nesterov1994}
Y.~Nesterov and A.~Nemirovski, \emph{Interior-Point Polynomial Algorithms in
  Convex Programming}, ser. SIAM Stud. Appl. Numer. Math.\hskip 1em plus 0.5em
  minus 0.4em\relax Philadelphia, PA, USA: SIAM, 1994, vol.~13.

\bibitem{BoVa09}
S.~Boyd and L.~Vandenberghe, \emph{Convex Optimization}.\hskip 1em plus 0.5em
  minus 0.4em\relax Cambridge, U.K.: Cambridge Univ. Press, 2004.

\bibitem{Matthiesen2018a}
B.~Matthiesen and E.~A. Jorswieck, ``Efficient global optimal resource
  allocation in non-orthogonal interference networks,'' \emph{{IEEE} Trans.
  Signal Process.}, 2019.

\bibitem{Tuy2009}
H.~Tuy, ``{$\mathcal{D(C)}$}-optimization and robust global optimization,''
  \emph{J. Global Optim.}, vol.~47, no.~3, pp. 485--501, 10 2009.

\bibitem{wcnc18}
B.~Matthiesen, Y.~Yang, and E.~A. Jorswieck, ``Optimization of weighted
  individual energy efficiencies in interference networks,'' in \emph{Proc.
  IEEE Wireless Commun. Netw. Conf. (WCNC)}, Barcelona, Spain, Apr. 2018.

\bibitem{Matthiesen2017}
B.~Matthiesen and E.~A. Jorswieck, ``Global sum rate optimal resource
  allocation for non-regenerative 3-way relay channels,'' in \emph{Workshop
  Broadband Wireless Commun. Comput. Boards (Atto-Nets), IEEE Int. Conf,
  Ubiquitous Wireless Broadband (ICUWB)}, Salamanca, Spain, Sep. 2017.

\bibitem{Luo2008}
Z.-Q. Luo and S.~Zhang, ``Dynamic spectrum management: Complexity and
  duality,'' \emph{{IEEE} J. Sel. Areas Commun.}, vol.~2, no.~1, pp. 57--73,
  Feb. 2008.

\bibitem{Knuth1997vol3}
D.~E. Knuth, \emph{The Art of Computer Programming: Sorting and Searching},
  2nd~ed.\hskip 1em plus 0.5em minus 0.4em\relax Reading, MA, USA:
  Addison-Wesley, 1997, vol.~3.

\bibitem{Schaible1993}
S.~Schaible, ``Fractional programming,'' in \emph{Handbook of Global
  Optimization}, ser. Nonconvex Optim. Appl., R.~Horst and P.~M. Pardalos,
  Eds.\hskip 1em plus 0.5em minus 0.4em\relax Norwell, MA, USA: Kluwer, 1995,
  vol.~2, pp. 495--608.

\bibitem{HaRa51}
O.~Hanner and H.~R{\aa}dstr{\"o}m, ``A generalization of a theorem of
  {F}enchel,'' \emph{Proc. Amer. Math. Soc.}, vol.~2, no.~4, pp. 589--593,
  1951.

\bibitem{BaShSh06}
M.~S. Bazaraa, H.~D. Sherali, and C.~M. Shetty, \emph{Nonlinear Programming:
  Theory and Algorithms}, 3rd~ed.\hskip 1em plus 0.5em minus 0.4em\relax
  Hoboken, NJ, USA: Wiley, 2006.

\bibitem{Ke60}
J.~E. Kelley, Jr., ``The cutting-plane method for solving convex programs,''
  \emph{J. Soc. Indust. Appl. Math.}, vol.~8, no.~4, pp. 703--712, 1960.

\bibitem{ElGamal2011}
A.~El~Gamal and Y.-H. Kim, \emph{Network Information Theory}.\hskip 1em plus
  0.5em minus 0.4em\relax Cambridge, U.K.: Cambridge Univ. Press, 2011.

\bibitem{MoZhCi06}
M.~Mohseni, R.~Zhang, and J.~Cioffi, ``Optimized transmission for fading
  multiple-access and broadcast channels with multiple antennas,'' \emph{{IEEE}
  J. Sel. Areas Commun.}, vol.~24, no.~8, pp. 1627--1639, Aug. 2006.

\bibitem{He17}
\BIBentryALTinterwordspacing
C.~Hellings, ``Reduced-entropy signals in {MIMO} communication systems,'' Ph.D.
  Thesis, Professur f{\"u}r Methoden der Signalverarbeitung, Technische
  Universit{\"a}t M{\"u}nchen, 2017. [Online]. Available:
  \url{http://mediatum.ub.tum.de/?id=1362885}
\BIBentrySTDinterwordspacing

\bibitem{Knuth1997vol1}
D.~E. Knuth, \emph{The Art of Computer Programming: Fundamental Algorithms},
  3rd~ed.\hskip 1em plus 0.5em minus 0.4em\relax Reading, MA, USA:
  Addison-Wesley, 1997, vol.~1.

\bibitem{Sun2017a}
Y.~Sun, P.~Babu, and D.~P. Palomar, ``Majorization-minimization algorithms in
  signal processing, communications, and machine learning,'' \emph{{IEEE}
  Trans. Signal Process.}, vol.~65, no.~3, pp. 794--816, Feb. 2017.

\bibitem{Schoebel2009}
A.~Sch{\"o}bel and D.~Scholz, ``The theoretical and empirical rate of
  convergence for geometric branch-and-bound methods,'' \emph{J. Global
  Optim.}, vol.~48, no.~3, pp. 473--495, Dec. 2009.

\end{thebibliography}
\end{document}